\documentclass[traditabstract]{aa}
\usepackage{txfonts}
\usepackage[english]{babel}
\usepackage{t1enc}
\usepackage[ansinew]{inputenc}
\usepackage{graphicx}
\usepackage{todonotes}
\usepackage{subfig}
\usepackage{lscape}
\usepackage{natbib}
\usepackage{footmisc}
\usepackage{url}
\usepackage{dingbat}
\usepackage{aalongtable} 
\begin{document}

\title{Recent star formation in the inner Galactic Bulge seen by ISOGAL
%\thanks{This is paper no. 2 in a refereed
%journal based on data from the ISOGAL project}\fnmsep
%\thanks{Based on observations with ISO, an ESA project with instruments
%funded by ESA Member States (especially the PI countries: France,
%Germany, the Netherlands and the United Kingdom) and with the
%participation of ISAS and NASA}
}
\subtitle{II -- The Central Molecular Zone}

\author{K. Immer\inst{1,2} \and F. Schuller\inst{1} \and A. Omont\inst{3} \and K. M. Menten\inst{1}}

\institute{
Max-Planck-Institut f\"ur Radioastronomie, Auf dem H\"ugel 69, D-53121 Bonn, Germany
\and
Harvard-Smithsonian Center for Astrophysics, 60 Garden Street, 02138 Cambridge, MA, USA
\and
Institut d'Astrophysique de Paris, CNRS, 98 bis, Bd Arago, F-75014 Paris, France
}
\date{Received 09/08/2011; Accepted 08/11/2011}
\email{kimmer@mpifr-bonn.mpg.de}

\abstract{We present 5--38 $\mu$m spectroscopic observations of a sample of 68 ISOGAL sources with unknown natures, taken with the Spitzer Infrared Spectrograph. Based on the characteristics and the slope of their spectra we classified the sources as young or late-type evolved objects. These sources were selected to test selection criteria based on the ISOGAL [7]--[15] color and the spatial extent parameter $\sigma_{\rm 15}$. We revised these criteria until they reliably distinguished between young and late-type evolved objects and then applied them to all ISOGAL sources in the central molecular zone (CMZ), resulting in the selection of 485 sources believed to be young. Furthermore, we added 656 Midcourse Space Experiment (MSX) sources to the CMZ sample that fulfilled $F_{\rm E}/F_{\rm D} > 2$ with $F{\rm D}$ and $F_{\rm E}$ being the flux densities in the D (15~$\mu$m) and E (21~$\mu$m) bands. After obtaining $\frac{L_{\rm bol}}{F_{\rm 15}}$ conversion factors, we calculated the bolometric luminosity, $L_{\rm bol}$, values for the CMZ sample and subsequently the masses of the sources. Applying a Kroupa initial mass function, we derived the total mass in young objects that has been formed over the last 1 Myr, resulting in an average star formation rate of 0.08 solar masses per year for the CMZ.}


\authorrunning{K. Immer et al.}
\titlerunning{Recent star formation in the inner Galactic Bulge II -- The CMZ}

\maketitle

\section{Introduction}

The central few hundred parsecs of the Milky Way present an exceptional environment in our Galaxy. This region produces around 5--10\% of the infrared luminosity of the Galaxy and contains ca. 10\% of its neutral gas ($\sim$5--10$\cdot$10$^7$ M$_{\sun}$) \citep[e.g.][]{Guesten1989, Rodriguez2004}. The conditions in this central molecular zone (CMZ, the inner $\pm$~1.5$\degr$~$\times$~$\pm$~0.5$\degr$ around the Galactic Center) are very extreme, the temperature, the velocity dispersion, the pressure and the estimated magnetic field strengths are much higher than in the Galactic disk \citep{Morris1996}. From a statistical point of view, a study of the star formation activity in the CMZ can uncover if star formation in this region is enhanced or suppressed by these extreme conditions.

The CMZ harbors many sites of massive star formation. The prominent emission of giant molecular clouds from infrared to radio wavelengths indicates recent bursts of star formation near the Galactic Center. However, most star formation studies in the CMZ are focused on the most massive and highly active star clusters near Sgr A$^{*}$ or the giant \ion{H}{II} complexes Sgr B2 and Sgr C.

Thus, in order to obtain an extensive overview of the star formation activity in the CMZ, a large number of young stellar objects has to be identified in this region. A method which is applicable on a large number of sources to identify YSO candidates is the usage of color selection criteria. However, in order to be reliable, photometric color criteria require always a spectroscopic confirmation of the nature of objects in a well selected sample.

At a particular time in its evolution a young stellar object (YSO) begins to heat its dust in the surrounding birth cloud which then re-emits this absorbed energy at infrared wavelengths. Studying massive star formation at infrared wavelengths (preferably with spectroscopic methods) gives therefore insight into the characteristics of the circumstellar dust as well as the properties of the embedded star. 

Large surveys combining several bands in the near- and mid-infrared are important tools to carry out systematic studies of young stellar populations. In this publication, we will use data of the ISOGAL \citep{Omont2003} and the Midcourse Space Experiment \citep[MSX,][]{Price2001} surveys. The ISOGAL survey comprises observations of the inner Galaxy at 7 and 15 $\mu$m, taken with the infrared camera ISOCAM on board the Infrared Space Observatory (ISO) with a spatial resolution better than 6$\arcsec$. The MSX project is a multi-wavelengths infrared survey of the entire Galactic plane in four spectral bands at 8, 12, 15, and 21 $\mu$m with a spatial resolution of $\sim$18$\arcsec$.

 \citet{Schuller2006} (hereafter: Paper I) defined selection criteria based on the [7]--[15] ISOGAL color (where [7] and [15] are the magnitudes at 7 and 15~$\mu$m, respectively, measured with ISO), as well as the magnitude uncertainty at 15~$\mu$m, $\sigma_{\rm 15}$ \citep[residual between the fitted point spread function and the actual source profile, see][]{Schuller2003}, which serves as a parameter for the spatial extension of the source (explained below). These criteria were established from a group of point sources from the ISOGAL point source catalogue \citep{Schuller2003, Omont2003} with known natures, dividing them into young objects\footnote{In this publication, the term `young object' refers to YSOs (a phase in the star formation process before the ionization of the surrounding birth cloud) and \ion{H}{II} regions (a later phase where the star starts to ionize its dust cocoon).} and late-type evolved objects (LEOs). Since young objects are embedded in a dust cocoon, they appear more extended at 15~$\mu$m than LEOs.  As explained in Paper I, a high value of $\sigma_{\rm 15}$ indicates that the source is marginally resolved at 15~$\mu$m. To be present in the PSC, such sources still have angular diameters $\la30\arcsec$ but values of 10$\arcsec$--20$\arcsec$ (FWHM) are typical.

In this publication, these selection criteria are tested on a sample of sources with unknown evolutionary stages. We will determine the nature of these sources from the characteristics and the slope of their mid-infrared spectra which were obtained with the Spitzer Infrared Spectrograph \citep{Houck2004}. By means of this sample of unknown sources we will refine the selection criteria until they reliably distinguish between young and late-type evolved objects.

The final goal is to apply a high quality version of the ISO selection criteria on all ISOGAL sources in the CMZ in order to find young objects and to characterize the star formation activity near the center of our Galaxy.

This paper is divided into two parts. The first part comprises the analysis of the infrared spectra. In Sections \ref{SourSel}, \ref{ObsIRS}, and \ref{ReducIRS} we present the selection of sources for our test sample, the Spitzer observations of the infrared spectra and the data processing of the spectra. In the following subsection, the classification of the studied sources is described. In the last subsection of this part, we will give refined selection criteria, which will then be applied to all ISOGAL sources in the CMZ in the second part of this publication in order to find young objects within the inner 3$\degr$ of our Galaxy. In Section \ref{DiscussionPartII} we derive the bolometric luminosities as well as the masses of all sources in our CMZ sample and finally determine the average star formation rate in the CMZ.

\section{Verification of the ISO color selection criteria}

\subsection{Source selection}
\label{SourSel}

Infrared colors are a powerful tool to distinguish between various classes of objects. Based on fluxes at 12, 25, and 60~$\mu$m measured with the Infrared Astronomy satellite (IRAS), \citet{Wood1989} defined criteria to identify ultra-compact \ion{H}{II} (UC\ion{H}{II}) regions based on far-IR colors. In the 1990s, mid-IR surveys conducted with MSX and ISO covered the Galactic plane with much higher spatial resolution and sensitivity than IRAS. \citet{Lumsden2002} defined color criteria based on MSX and the near-IR 2MASS survey to select candidate massive YSOs. Similarly, from identifications of ISOGAL point sources with known objects, it has been found that massive YSOs and UC\ion{H}{II} regions show [7]--[15] colors greater than 2~mag \citep{Felli2002}. Moreover, most sources with [15]~$<$~4~mag and [7]--[15]~$>$~2~mag have MSX counterparts with flux ratios F$_{21~\mu{\rm m}}$/F$_{15~\mu{\rm m}}$ above 2, making their interpretation as YSO very likely, since YSOs are characterized by strong rising continuum emission in the IR range. In addition, in Paper I it has been shown that the spatial extent at 15~$\mu$m, as it appears in the ISOGAL catalogue, can help in determining whether sources brighter than 300~mJy at 15~$\mu$m (or [15]~$\leq$~4.5~mag) are young objects or late-type evolved stars. Infrared sources with ISOGAL characteristics similar to those of known massive young objects, but without any known counterpart, are thus good massive YSO candidates. 

Not all of the sources in the ISOGAL catalogue have been detected at both of the covered wavelengths (7 and 15~$\mu$m). Since sources brighter than 8~mag at 7~$\mu$m should have been detected by the ISOGAL survey, a lower limit for the [7] magnitude of 8~mag was assigned to sources which were not detected at 7~$\mu$m.

Near the Galactic Center, \citet{Glass2001} published a catalogue of long period variable stars, which also appear as bright ISOGAL sources. In the 0.1~deg$^2$ region of the ISOGAL survey overlapped by the \citet{Glass2001} observations, 95 sources brighter than 5.25~mag at 15~$\mu$m show [7]--[15] colors larger than 1.8~mag. Sixteen of these sources can be matched with long period variable stars from the \citet{Glass2001} catalogue, and were removed from our sample. After rejecting another six sources associated with known OH/IR stars and five showing near-IR spectra typical of AGB stars \citep{Schultheis2003}, we finally selected 68 ISOGAL sources as targets for follow-up spectroscopic observations with Spitzer. This test sample is complete down to a 15~$\mu$m flux density of 140~mJy over a 0.1 sq. deg. area.

Despite our effort to remove known LEOs from our test sample, we realized later that two of our targets (J174602.8--290359, J174623.0--285845) have been identified as OH/IR stars by \citet{Sjouwerman1998}. These sources have been helpful in establishing the identification criteria for LEOs in our test sample.

\subsection{Observation and Inspection of the Data}
\label{ObsIRS}

\begin{figure}
	\centering
		\includegraphics[angle=90, width=9cm]{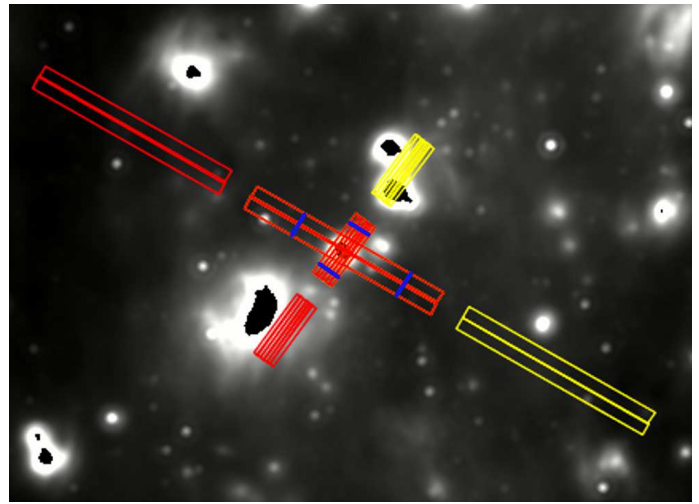}
	\caption{Footprint of the IRS slits shown on a 24~$\mu$m image around J174505.6--291018. In that case, the map consists of 6 pointings with the SL module, and 2 pointings with the LL module. Positions in the slits where background spectra were obtained are shown in blue.}
	\label{J174505.6-24um}
\end{figure}

The spectroscopic data were obtained in March 2005 with the two low resolution modules of the Infrared Spectrograph \citep[IRS;][]{Houck2004}, onboard the Spitzer Space Telescope \citep{Werner2004}. The 68 sources were observed in the IRS-mapping mode, with the low-resolution modules SL (Short-wavelength, Low-resolution) and LL (Long-wavelength, Low-resolution). Both modules are subdivided in three submodules that cover different parts of the two wavelength ranges (SL2: 5--7.5~$\mu$m, SL3: 7.3--8.7~$\mu$m, SL1: 7.5--15~$\mu$m, LL2: 14--21~$\mu$m, LL3: 19.4--21.7~$\mu$m, LL1: 21--40~$\mu$m). Each map consisted of between 3 and 17 pointings with the SL module, and between 1 and 4 pointing(s) with the LL module, depending on the size of the source as seen in the ISOGAL image. As an example, Fig. \ref{J174505.6-24um} shows a footprint of the IRS slits on a Spitzer/MIPS 24~$\mu$m image around the source J174505.6--291018. The ramp duration was set to 6~s for each module, and two cycles were required per ramp. In most cases, this was sufficient to achieve S/N ratios of 100 or more over the entire wavelength range.\\ 

We began with a first inspection of the data. Depending on the detection of the ISOGAL sources at short (5--15~$\mu$m) or long (15--40~$\mu$m) mid-infrared wavelengths in the spectroscopic data, we divided the sources into different groups:
\begin{itemize}
	\item A) ISOGAL sources, detected as one source in the SL as well as in the LL data at the same position (49 sources)
	\item B) ISOGAL sources, only detected in the LL data (8 sources)
	\item C) ISOGAL sources, detected neither in the SL nor in the LL data (4 sources)
	\item D) Confusion of several Spitzer sources in the SL or LL data at the position of the ISOGAL source (7 sources)
\end{itemize}

Only the sources in groups A and B were processed further, giving a total of 57 sources. Since the group C ISOGAL sources are not detected with Spitzer at higher resolution, these sources are probably false detections in the ISOGAL catalogue.

\subsection{Data reduction}
\label{ReducIRS}

The data reduction was conducted with the Spitzer software IRSCLEAN\footnote{\scriptsize{\url{http://ssc.spitzer.caltech.edu/archanaly/contributed/irsclean/IRSCLEAN\_MASK.html}}} (to clean rogue pixels in the data) and SPICE\footnote{\scriptsize{\url{http://ssc.spitzer.caltech.edu/dataanalysistools/tools/spice/}}} (to extract the spectra). The spectra are calibrated by SPICE with calibration files which are automatically selected by the software (for more information about the calibration see the SPICE User's Guide\footnote{\scriptsize{\url{http://ssc.spitzer.caltech.edu/postbcd/doc/spice.pdf}}}).

A direct background subtraction cannot be conducted with SPICE. Therefore, the source and the background spectra were determined separately and the background subtraction was conducted later. Since the infrared background in the Galactic Center region varies greatly on short spatial scales, the background spectra were obtained from the same subslit as the source spectra. Several background spectra were extracted from different positions and pointings around the source (the number of obtained background spectra varies from source to source, depending on the infrared source density around the target sources). The blue lines in Fig. \ref{J174505.6-24um} show the positions where background spectra were obtained around J174505.6--291018. In order to obtain a mean background spectrum at the position of the source, the weighted average of the background spectra and the associated errors were determined.

Per subslit and submodule (SL2, SL3, SL1, LL2, LL3, LL1), the source spectra were extracted from those pointings where the emission of the source could be distinguished from the background emission and the emission of other infrared sources. Per pointing, the source spectra of the two cycles were averaged and the mean background spectrum was subtracted. Per submodule, all background subtracted source spectra were added together. Then, the background subtracted source spectra of the six submodules were assembled to produce the final source spectrum.

Seventeen sources of the test sample show flux jumps between the SL and the LL module spectra around 14~$\mu$m. These are due to extended emission that falls in the slits in combination with the different slit widths of the two modules covering areas of different sizes around the sources.

Twentyseven of the 57 test sample sources were only detected at 15~$\mu$m in ISOGAL and a lower limit of 8~mag was assigned to their [7] magnitudes. Therefore, their [7]--[15] colors are also only lower limits. Furthermore, it should be mentioned that the ISOGAL magnitudes at 7 and 15~$\mu$m were not observed at the same time for 63\% of the 57 sources (sources with right ascension, RA,~<~17:45:40). Due to possible variability of the sources, this could lead to errors in the [7]--[15] parameter.

\subsection{Spectroscopic Properties of the Test Sample}

The most common features in the IRS wavelength range, detected in our sources, are silicate absorption bands, emission features from polycyclic aromatic hydrocarbons, and fine structure lines from ionized species. A broad silicate absorption band at around 9.8~$\mu$m is present in all the spectra but the central wavelength changes marginally from source to source. Further silicate absorption features are detected at 18, 22.3 and/or 23~$\mu$m \citep[e.g.][]{Molster2002}. These silicate features are indicative of O-rich envelopes with large mass infall rates.

In a few sources, absorption bands at around 5.95, 6.7, and 7.4~$\mu$m seem to be present. According to \citet{Gerakines1995}, the absorption at 6~$\mu$m is partly due to a bending mode of water ice. The 6.7~$\mu$m absorption band can partly be explained by absorption through methanol ice \citep{Quirico2000} whereas the band at 7.4~$\mu$m is possibly caused by acetaldehyde (CH$_3$HCO) ice \citep{Schutte1998}. 

Emission from Polycyclic Aromatic Hydrocarbons (PAHs) requires a strong ultraviolet field to be excited and thus is frequently found at the edges of \ion{H}{II} regions. Eleven of the sources show unambiguous emission of PAHs at one or more of the following wavelengths: 6.2~$\mu$m, 7.7~$\mu$m, 8.6~$\mu$m, 11.3~$\mu$m, and 16.4~$\mu$m. However, due to confusion of the ice absorption bands at wavelengths shorter than 8~$\mu$m, in several sources the PAH emission features in this wavelength range could not be unambiguously identified. 

In our source spectra, the detection of PAH emission bands is always seen in combination with at least one of the following forbidden fine structure lines: [\ion{Ar}{II}] (6.97~$\mu$m), [\ion{Ne}{II}] (12.83~$\mu$m) and [\ion{S}{III}] (18.7~$\mu$m). The most frequent emission line, [\ion{Ne}{II}], is detected in the direction of 21 sources. In five of the sources the detection is weak and could be spurious. The [\ion{Ar}{II}] and the [\ion{S}{III}] lines are observed in the direction of seven and fourteen sources, respectively. Although the occurrence of these emission lines often indicates the existence of \ion{H}{II} regions and therefore the presence of hot massive stars, this emission can also arise from the warm, ionized, interstellar medium surrounding the sources. This fact complicates the background subtraction. The detection of the emission lines should therefore be considered with caution. Nevertheless, the lines can still be used as a diagnostic tool.

In several sources, the S(1) and S(3) rotational emission lines of H$_2$ at 17.03~$\mu$m and 9.66~$\mu$m (within the broad silicate absorption feature) are detected which are tracers for warm molecular gas \citep[e.g.][]{Kaufman2006, Fiolet2010}.

\scriptsize
\begin{longtable}{ccccccccc}
\caption{\label{PAHIce} Detected spectral features of the sources. The first column contains the source names. Columns 2 through 8 give the different detected spectral features: PAH emission, emission of the forbidden fine structure lines [ArII], [NeII], and [SIII], emission of the rotational lines S(1) and S(3) of H$_2$, and silicate absorption at 22.3 and/or 23~$\mu$m. The last column lists the detection of ice absorption bands in the spectra. An unclear identification of a spectral feature was indicated with a question mark.}\\ \hline                      
Source name	&	PAH 	   &	[\ion{Ar}{II}]	&	[\ion{Ne}{II}]	&	[\ion{S}{III}]	&	\multicolumn{2}{c}{H$_2$} 	&	Absorption	&	Ice Absorption Bands\\
            & Emission & 6.97~$\mu$m & 12.83~$\mu$m & 18.7~$\mu$m & S(1) (17.03~$\mu$m) & S(3) (9.66~$\mu$m) & 22.3 and/or 23~$\mu$m &\\ \hline
\endfirsthead 
\caption{continued.}\\	\hline	 
Source name	&	PAH 	   &	[\ion{Ar}{II}]	&	[\ion{Ne}{II}]	&	[\ion{S}{III}]	&	\multicolumn{2}{c}{H$_2$} 	&	Absorption	&	Ice Absorption Bands\\
            & Emission & 6.97~$\mu$m & 12.83~$\mu$m & 18.7~$\mu$m  & S(1) (17.03~$\mu$m)  & S(3) (9.66~$\mu$m) & 22.3 and/or 23~$\mu$m &\\ \hline
\endhead \hline
\endfoot 
J174457.4--291003	&			&			&			&			& \checkmark?	&			&			&\checkmark?\\
J174459.1--290653	&			&			&			&			& 			&			& \checkmark	&\\	
J174500.7--291007	& \checkmark ?	& \checkmark ?	& \checkmark ?	&			&			& \checkmark	&			&\\
J174501.0--285622	&			&			&			& \checkmark	& \checkmark	&			& \checkmark	&\\
J174504.9--291146	&			& \checkmark	& \checkmark	& \checkmark ?	&			& \checkmark	&			&\\
J174505.1--290937	& \checkmark ?	& \checkmark ?	&			&			& \checkmark	&			& \checkmark	&\\
J174505.6--291018	& \checkmark	& \checkmark	& \checkmark	& \checkmark	&			&			&			&\checkmark?\\
J174506.5--291118	&			&			& \checkmark	& \checkmark	&			&			&			&\\  
J174508.0--290655	&			&			&			&			&			&			& \checkmark	&\\
J174508.0--291039	&			& \checkmark?	&			& \checkmark	&			&			&			&\\
J174508.1--290840	& \checkmark	&			& \checkmark	& \checkmark	& \checkmark?	&			& \checkmark ?	&\checkmark?\\
J174508.7--290348	& \checkmark	& \checkmark?	&			&			&			& \checkmark	&			&\\
J174511.3--290621	& \checkmark	& \checkmark ?	& \checkmark ?	&			&			&			&			&\\
J174512.6--290646	& \checkmark ?	&			&			&			&			&			& \checkmark	&\checkmark?\\
J174515.4--291213	&			&			&			&			&			&			&			&\\
J174515.9--290155	& \checkmark ?	& \checkmark ?	&			&			&			&			& \checkmark ?	&\checkmark?\\
J174516.2--290315	&			&			&			&			&			&			&			&\checkmark?\\
J174517.1--291155	&			& \checkmark ?	&			&			&			&			& \checkmark ?	&\\
J174517.8--290813	&			&			&			&			&			&			& \checkmark	&\\
J174518.1--290439	&			&			& \checkmark	& \checkmark	&			&			&			&\\
J174518.1--291051	& \checkmark ?	& \checkmark	& \checkmark	& \checkmark	&			&			&			&\\
J174518.3--290628	&			&			&			&			& \checkmark?	&			& \checkmark	&\checkmark?\\
J174520.0--290750 	&			&			&			&			& \checkmark	&			&  \checkmark	&\\
J174520.1--290638	&			&			&			&			&			&			&			& \\
J174520.7--290213	&			&			& \checkmark	& \checkmark	&			&			&			&\\
J174522.1--291059	& \checkmark ?	& \checkmark ?	& \checkmark	& \checkmark ?	&			& \checkmark?	&			&\\
J174522.4--290242	& \checkmark	& \checkmark ?	& \checkmark ?	&			& \checkmark?	&			& \checkmark	&\\
J174523.3--290331	&			&			&			&			&			& \checkmark	& \checkmark ?	&\\
J174523.5--290225	& \checkmark	&			& \checkmark	&			&			&			& \checkmark	&\\
J174523.9--290310	&			&			&			&			&			&			&			&\\
J174524.8--290529	&			&			&			&			& \checkmark	&			&			&\\
J174524.9--290318	&			& \checkmark ?	& \checkmark	&			&			&			&			&\\
J174525.7--290942	& \checkmark	& \checkmark	& \checkmark	& \checkmark	& \checkmark	&			&			&\\
J174527.2--291126	&			&			&			&			&			&			& \checkmark ?	&\\
J174527.4--290655	&			&			&			&			& \checkmark	&			& \checkmark ?	&\\
J174529.5--291021	& \checkmark ?	& \checkmark	& \checkmark	&			&			&			&			&\checkmark?\\
J174553.3--290406	&			&			&			& \checkmark ?	&			&			&			&\\
J174555.6--290521	& \checkmark ?	&			& \checkmark ?	&			&			& \checkmark?	&			&\\
J174557.7--290302	& \checkmark ?	&			&			&			&			& \checkmark?	& \checkmark	&\\
J174600.1--290150	&			&			& \checkmark	& \checkmark	&			&			& \checkmark	&\\
J174602.5--290427	&			&			&			&			& \checkmark?	&			&			&\\
J174602.8--290359	&			&			&			&			&			&			&			&\\
J174612.7--285958	&			&			& \checkmark	& \checkmark	&			&			& \checkmark	&\\
J174613.1--290327	& \checkmark ?	&			&			&			&			&			& \checkmark	&\\
J174613.5--290024	& \checkmark	&			&			& \checkmark	&			&			& \checkmark	&\\
J174617.6--285857	& \checkmark ?	&			&			&			&			&			& \checkmark	&\\
J174618.0--290245	& \checkmark	&			& \checkmark ?	& \checkmark ?	&			& \checkmark?	& \checkmark	&\\
J174618.2--290136	& \checkmark ?	&			&			&			&			&			&			&\\
J174619.4--285611	&			&			&			&			&			&			& \checkmark	&\\
J174620.8--290532	&			&			&			&			&			&			& \checkmark	&\\
J174623.0--285845	&			&			&			&			&			&			&			&\\
J174623.8--285300	&			&			&			&			&			&			&			&\\
J174624.9--290345	&			&			&			&			&			&			& \checkmark	&\\
J174631.1--290652	& \checkmark ?	&			& \checkmark	&			&			&			& \checkmark	&\\
J174631.2--285448	&			&			&			& \checkmark ?	&			&			& \checkmark	&\\
J174631.8--285027	& \checkmark	&			&			& \checkmark	&			&			&  \checkmark	&\\
J174632.9--285115	& \checkmark	& \checkmark	& \checkmark	& \checkmark	&			&			& \checkmark	&
\end{longtable} 

\twocolumn
\normalsize

Table \ref{PAHIce} shows the spectral features that were identified in the source spectra. The first column contains the source names. Columns 2--5 list the detection of PAH emission features as well as forbidden fine structure lines ([\ion{Ar}{II}], [\ion{Ne}{II}], [\ion{S}{III}]). Where the identification of a feature was not clear, question marks were set in the table. Columns 6--8 mark the detection of emission of the S(1) and S(3) rotational lines of H$_2$ at 17.03~$\mu$m and 9.66~$\mu$m as well as silicate absorption at 22.3 or 23~$\mu$m. In the last column, the detection of ice absorption bands at around 5.95, 6.7, or 7.4~$\mu$m are mentioned.

\begin{figure}
	\centering
	\includegraphics[width=9cm]{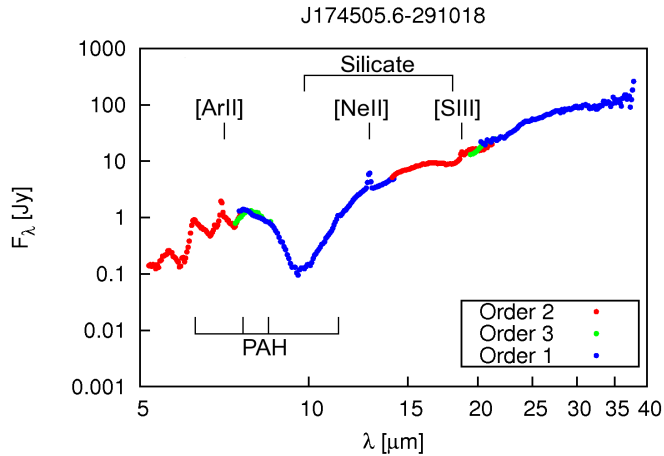}	
\caption{Background subtracted spectrum of the source J174505.6--291018. The data of the different subslits are presented by the three different colors (green: SL2 or LL2, blue: SL3 or LL3, red: SL1 or LL1). Denoted are the silicate absorption features, the PAH emission bands and the forbidden fine structure lines [\ion{Ar}{II}], [\ion{Ne}{II}], and [\ion{S}{III}].}
	\label{Spectrumr10457600}
	\end{figure}

As an example, Figure \ref{Spectrumr10457600} shows the background-subtracted spectrum of the source J174505.6--291018. The three different colors represent the data of the different subslits (green: SL2 or LL2, blue: SL3 or LL3, red: SL1 or LL1). The spectrum shows emission lines of [\ion{Ar}{II}], [\ion{Ne}{II}], and [\ion{S}{III}] as well as two silicate absorption features at 9.8 and 18~$\mu$m. Furthermore, the detected PAH emission features are indicated. The spectra of the remaining 56 sources are shown in Fig. \ref{Spectra} (online material).

\onlfig{3}{
%\begin{subfigures}
%\setcounter{figure}{1}
\begin{figure*}
	\caption{Background subtracted spectra. The data of the different subslits are presented by the three different colors (green: SL2 or LL2, blue: SL3 or LL3, red: SL1 or LL1).}
	\centering
	\subfloat{\includegraphics[width=7.5cm]{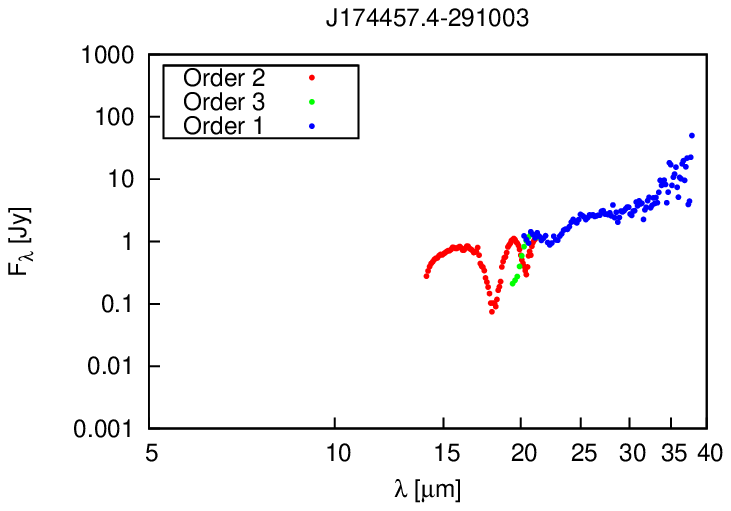}}	
	\subfloat{\includegraphics[width=7.5cm]{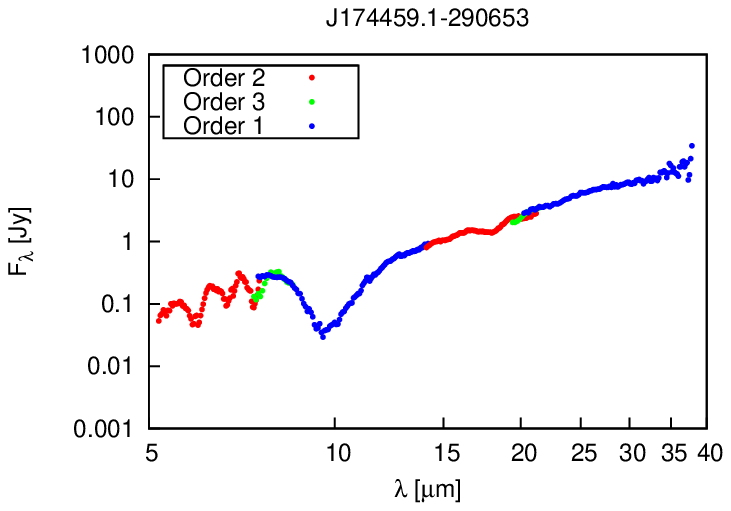}}\\
	\subfloat{\includegraphics[width=7.5cm]{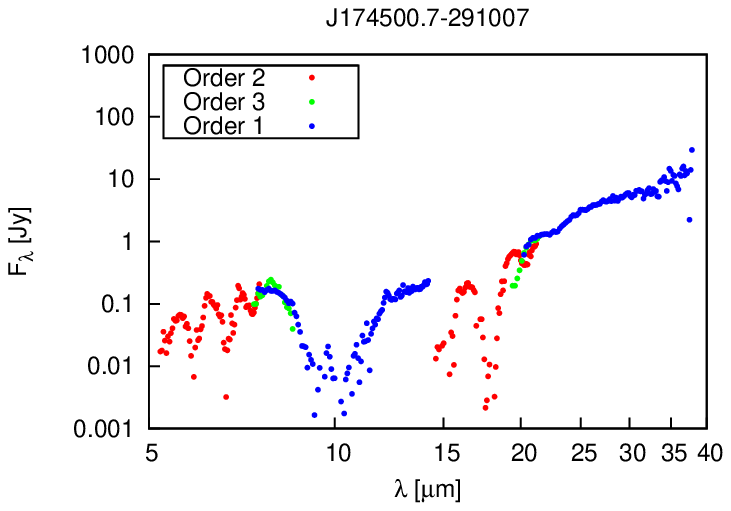}}
	\subfloat{\includegraphics[width=7.5cm]{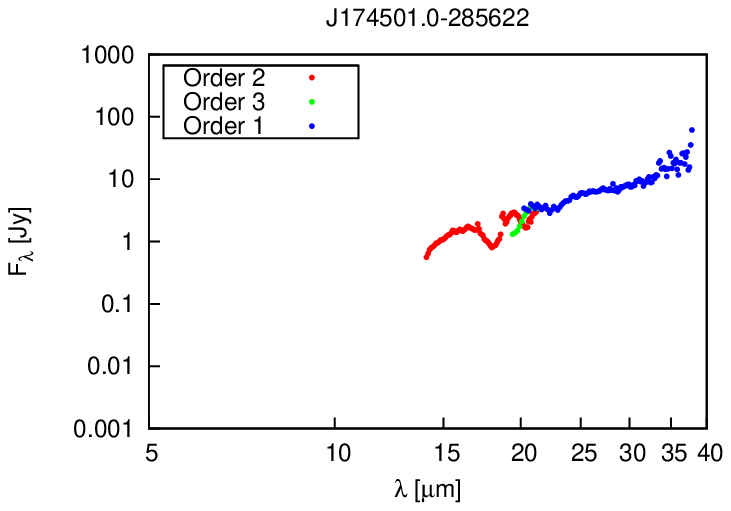}}\\	
	\subfloat{\includegraphics[width=7.5cm]{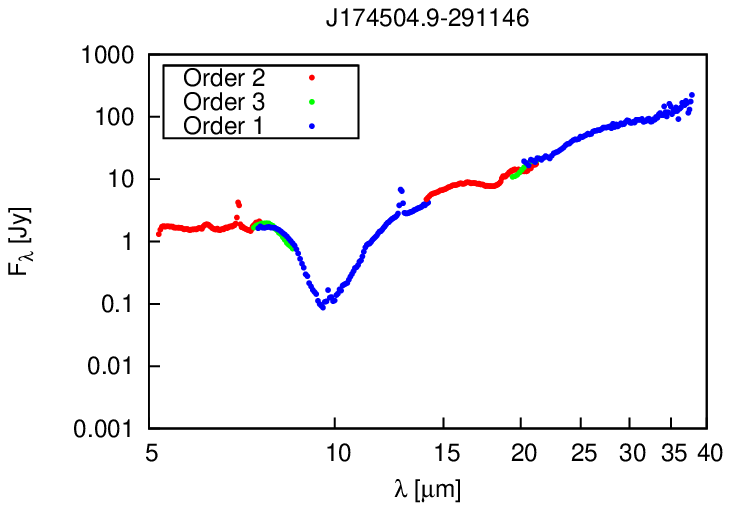}}
	\subfloat{\includegraphics[width=7.5cm]{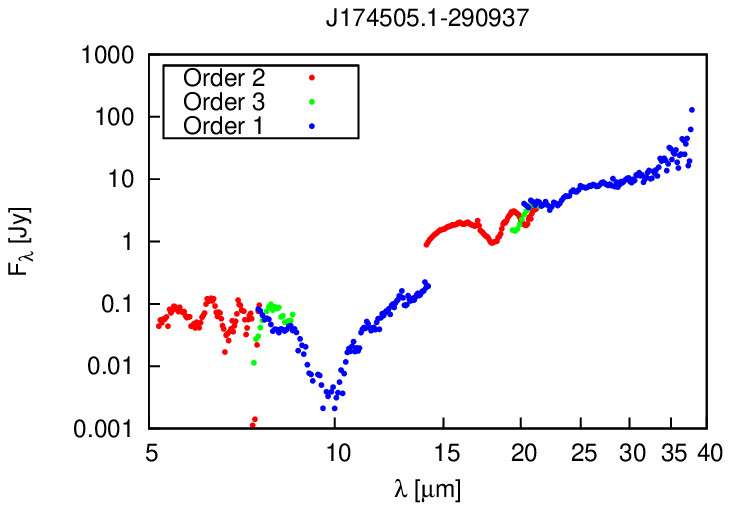}}\\	
	\subfloat{\includegraphics[width=7.5cm]{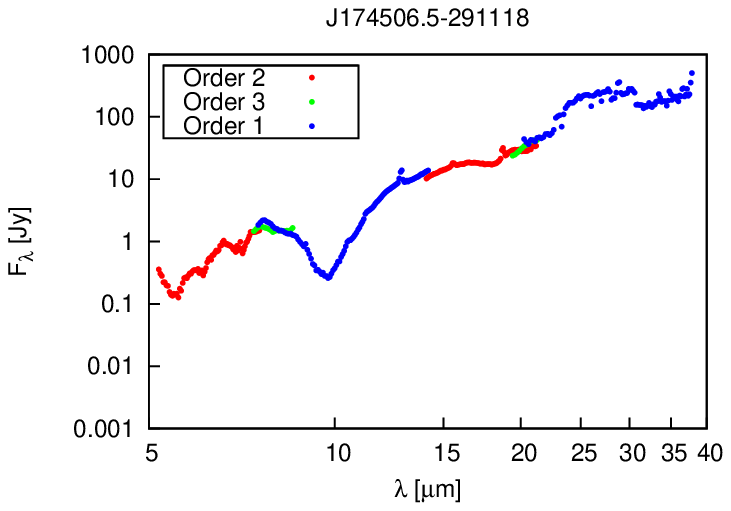}}
	\subfloat{\includegraphics[width=7.5cm]{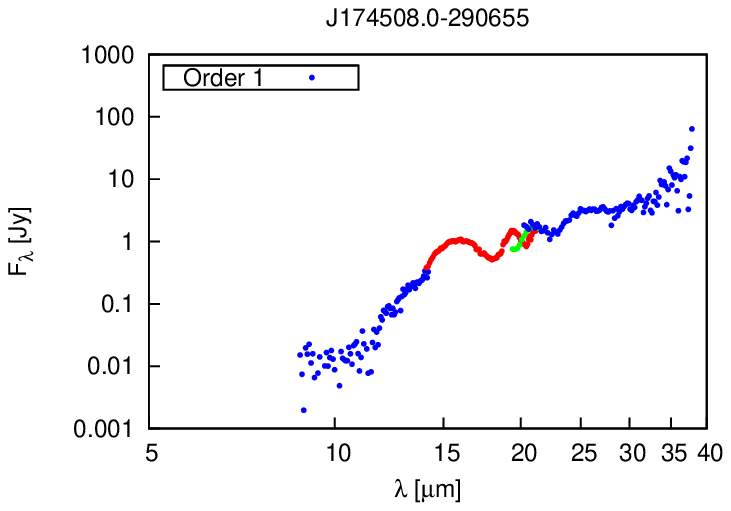}}
	\label{Spectra}
\end{figure*}

\addtocounter{figure}{-1}
\begin{figure*}
	\centering
	\caption{Continued}
	\subfloat{\includegraphics[width=7.5cm]{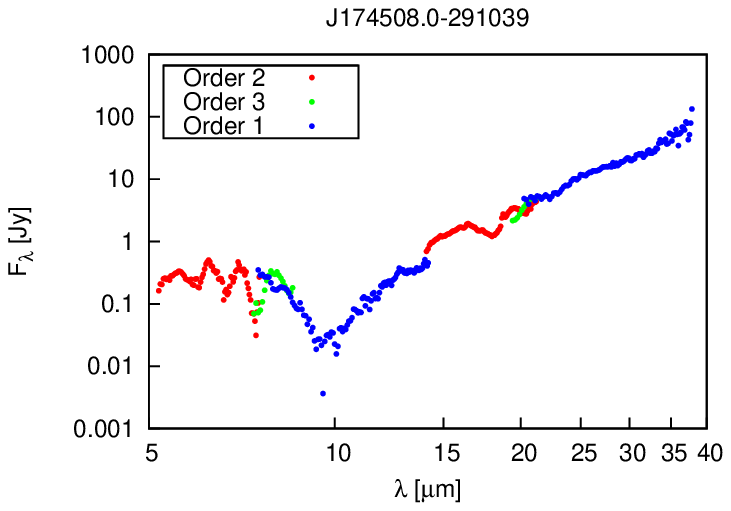}}
	\subfloat{\includegraphics[width=7.5cm]{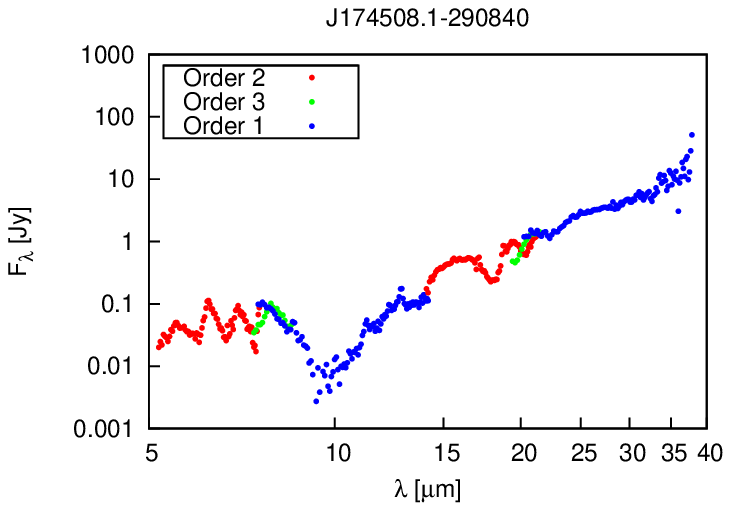}}\\
	\subfloat{\includegraphics[width=7.5cm]{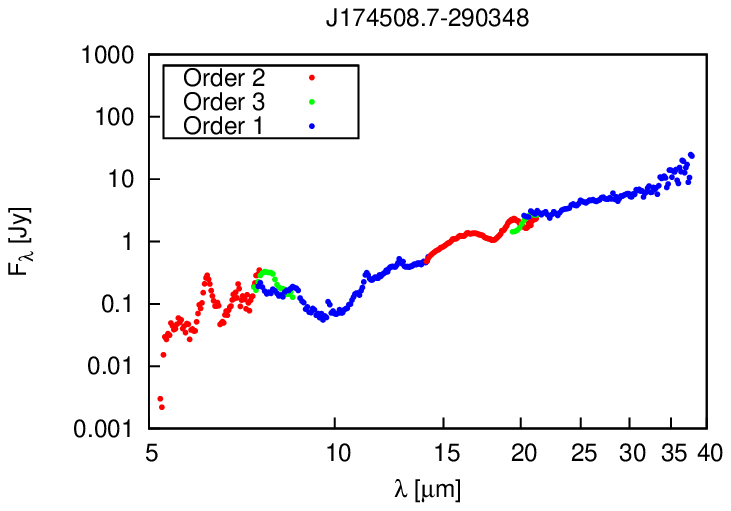}}
	\subfloat{\includegraphics[width=7.5cm]{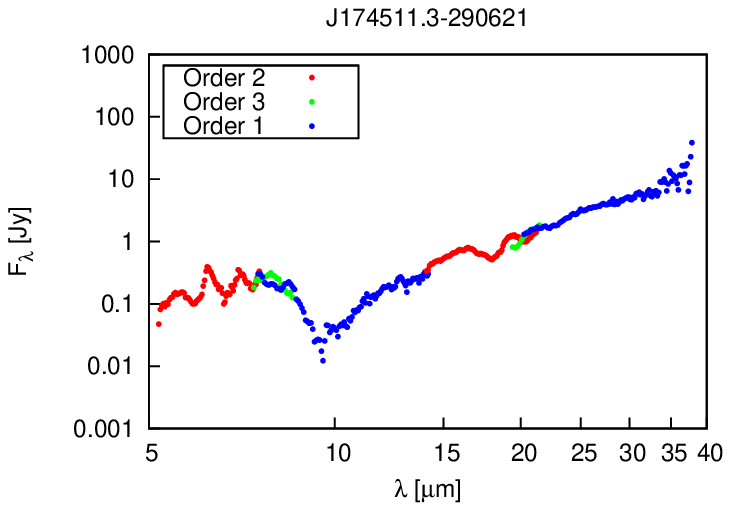}}\\
	\subfloat{\includegraphics[width=7.5cm]{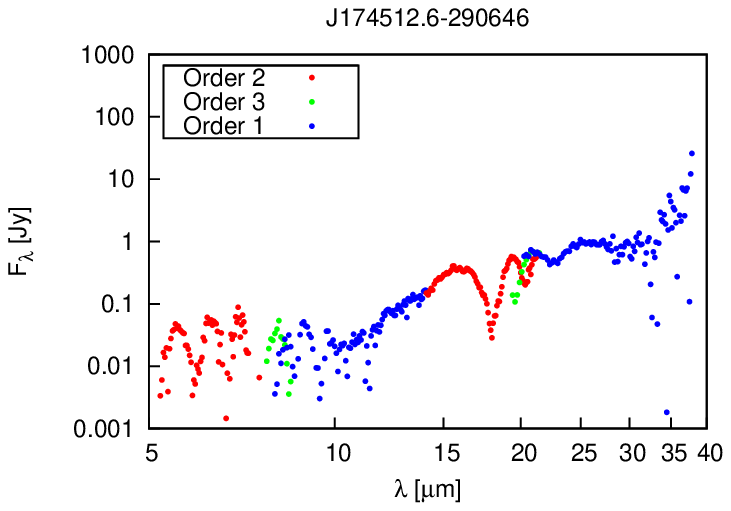}}
	\subfloat{\includegraphics[width=7.5cm]{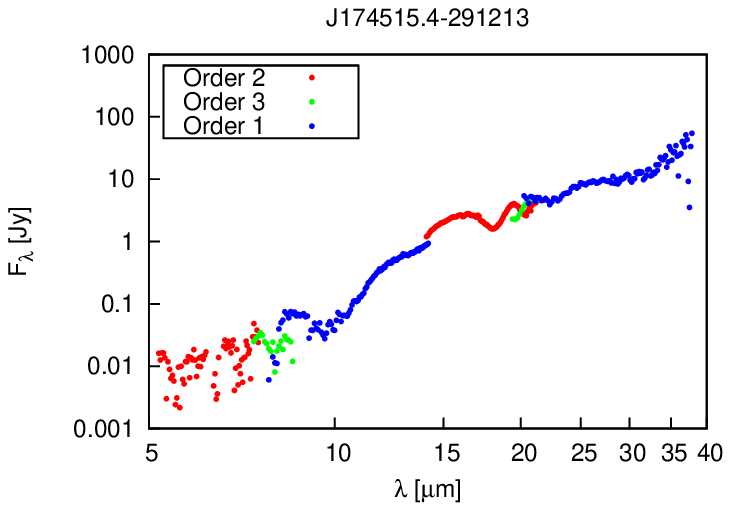}}\\
	\subfloat{\includegraphics[width=7.5cm]{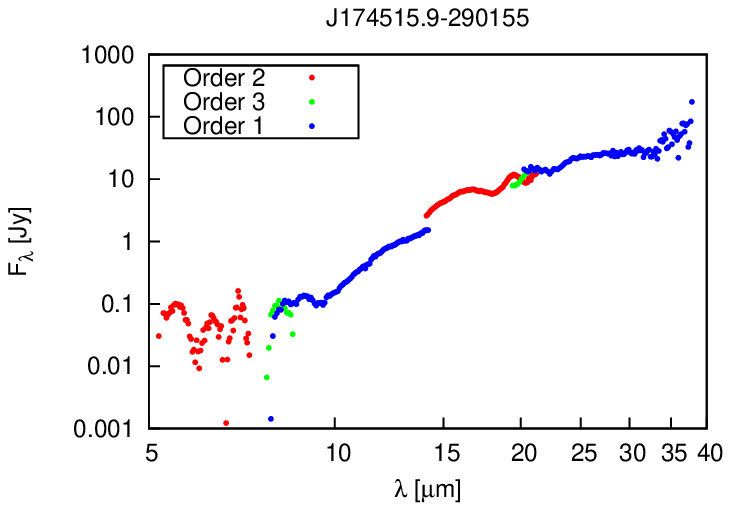}}
	\subfloat{\includegraphics[width=7.5cm]{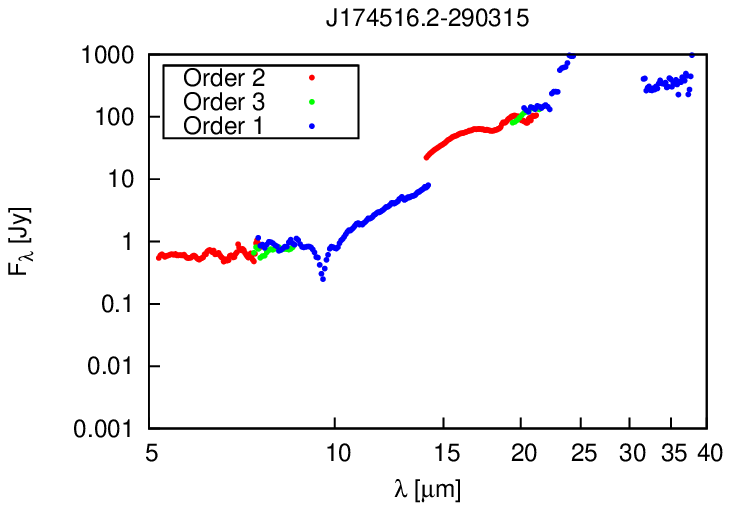}}	
\end{figure*}

\addtocounter{figure}{-1}
\begin{figure*}
	\centering	
	\caption{Continued}
	\subfloat{\includegraphics[width=7.5cm]{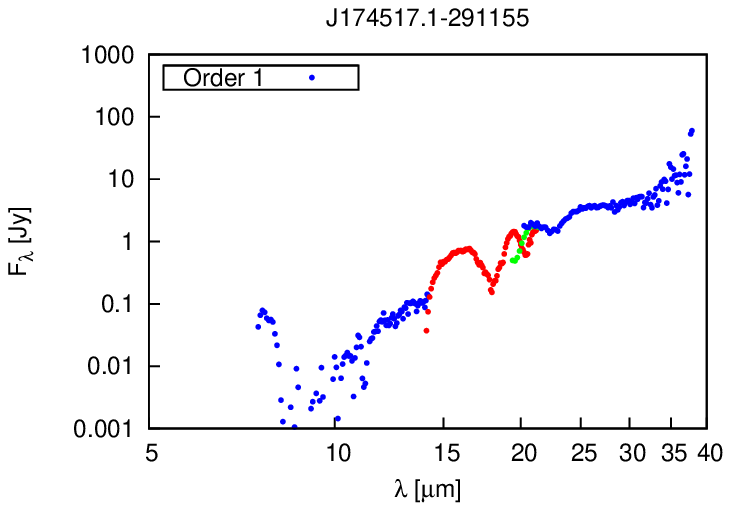}}
	\subfloat{\includegraphics[width=7.5cm]{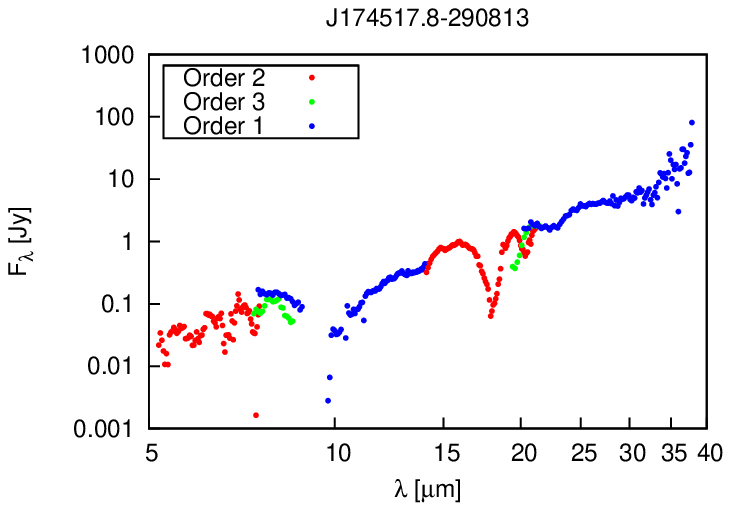}}\\
	\subfloat{\includegraphics[width=7.5cm]{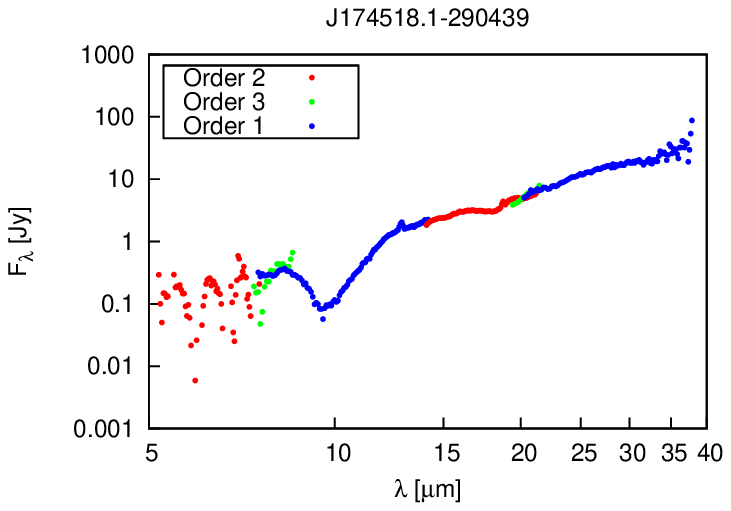}}	
	\subfloat{\includegraphics[width=7.5cm]{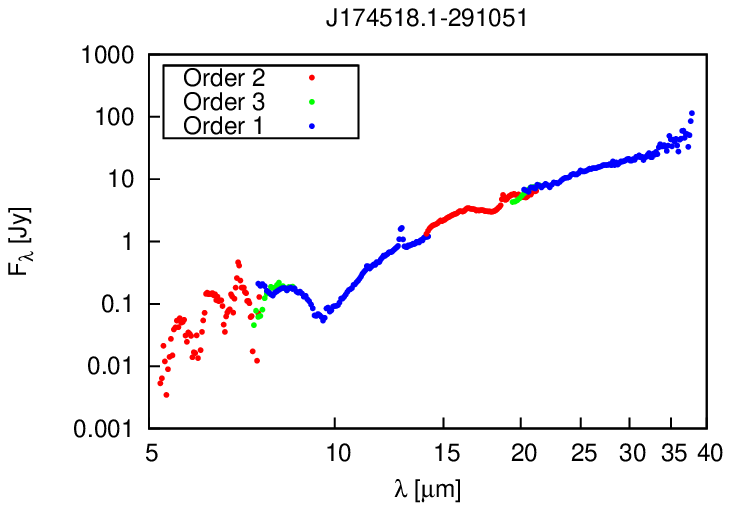}}\\
	\subfloat{\includegraphics[width=7.5cm]{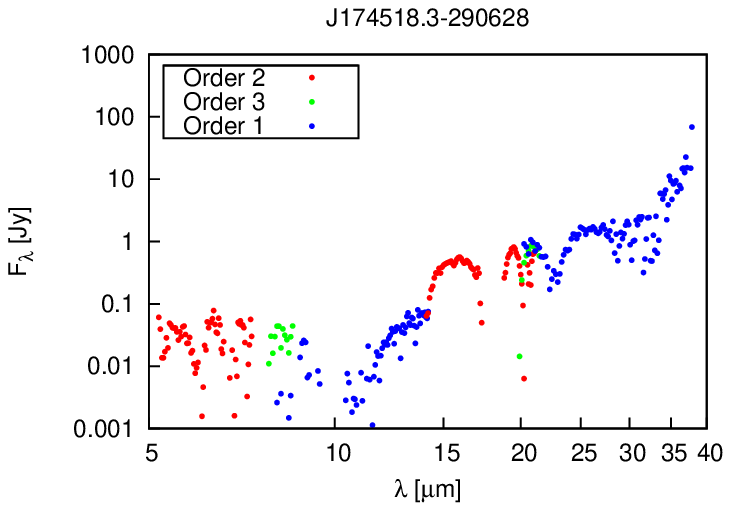}}
	\subfloat{\includegraphics[width=7.5cm]{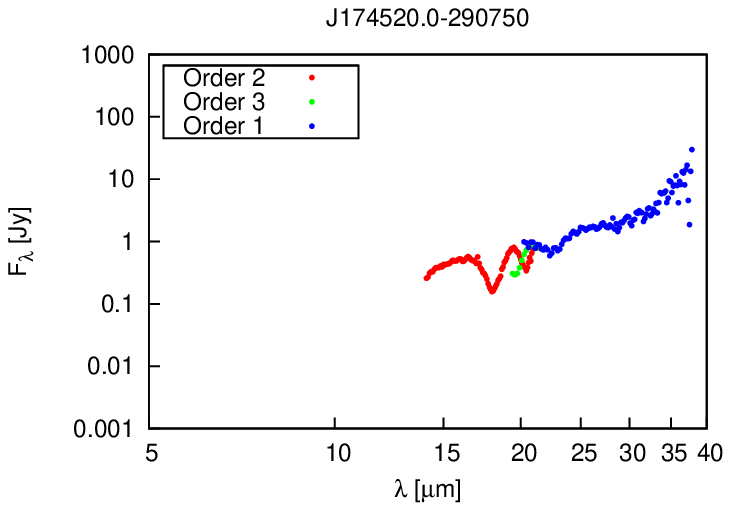}}\\
	\subfloat{\includegraphics[width=7.5cm]{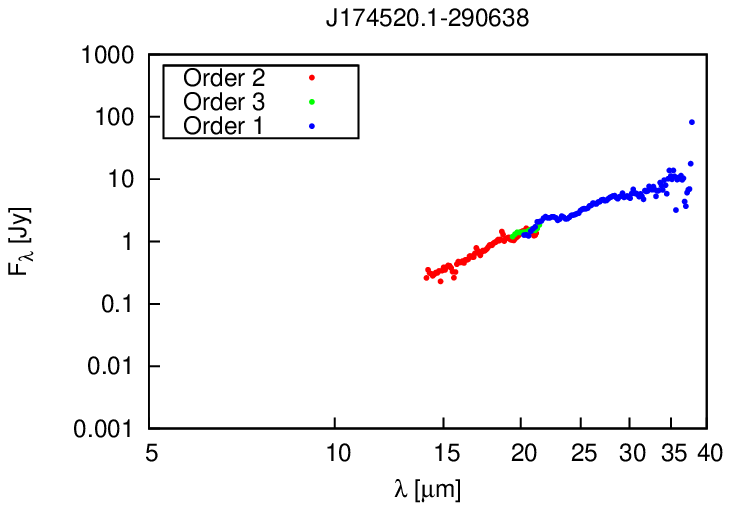}}
	\subfloat{\includegraphics[width=7.5cm]{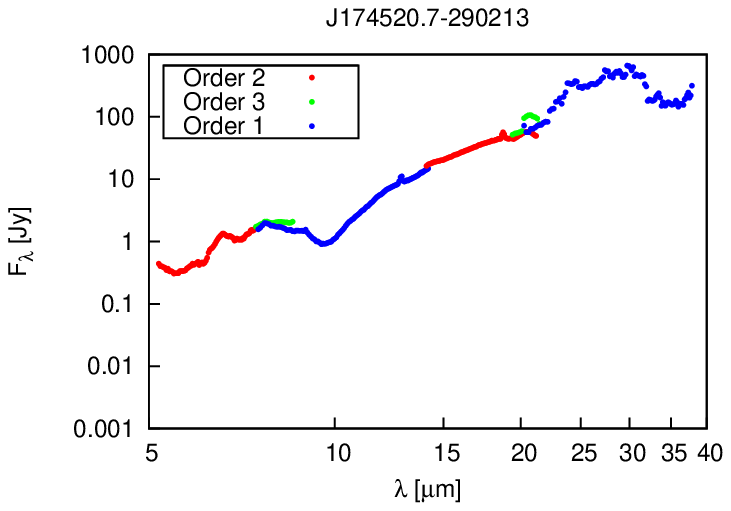}}
\end{figure*}

\addtocounter{figure}{-1}
\begin{figure*}
	\centering
	\caption{Continued}
	\subfloat{\includegraphics[width=7.5cm]{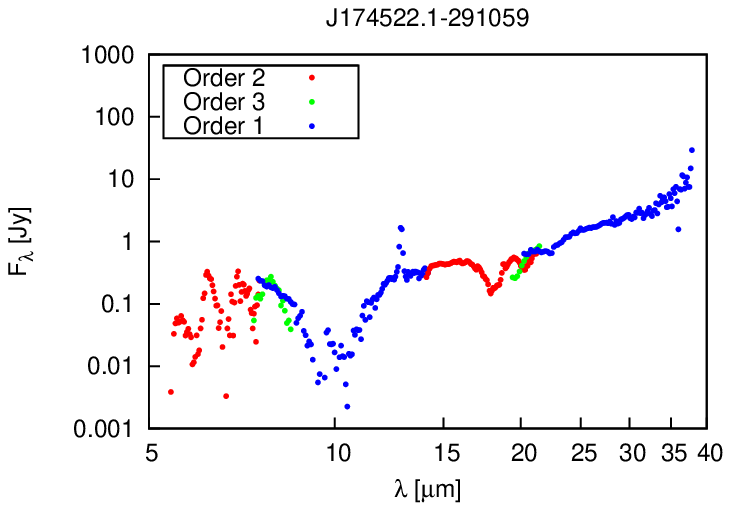}}	
	\subfloat{\includegraphics[width=7.5cm]{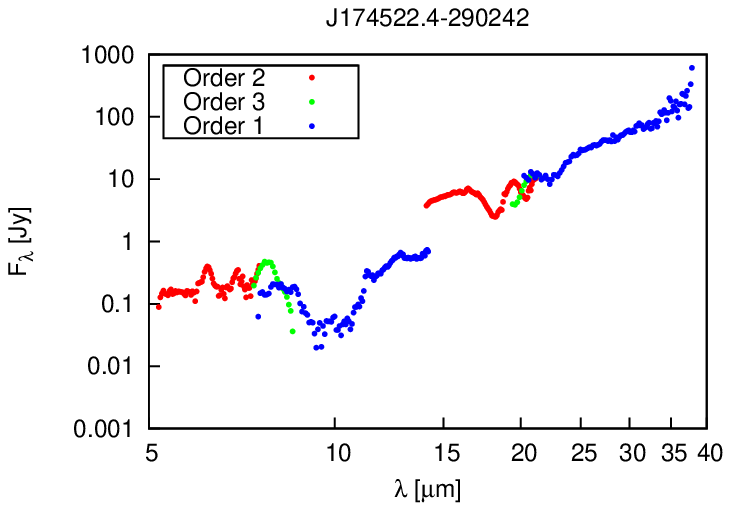}}\\
	\subfloat{\includegraphics[width=7.5cm]{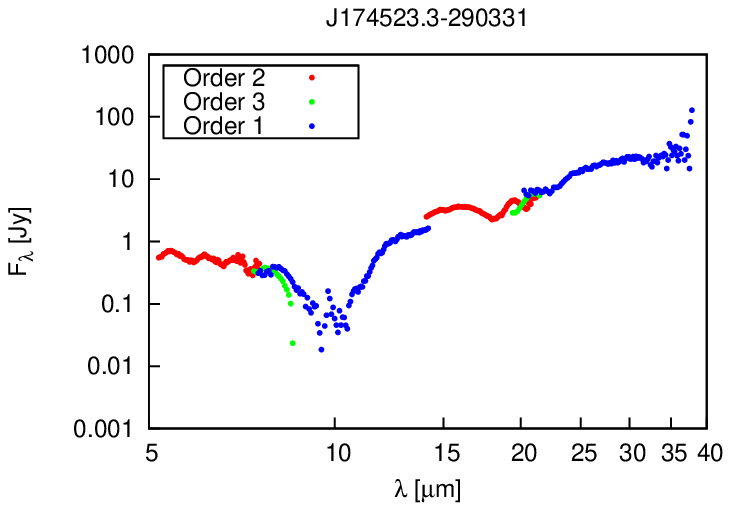}}
	\subfloat{\includegraphics[width=7.5cm]{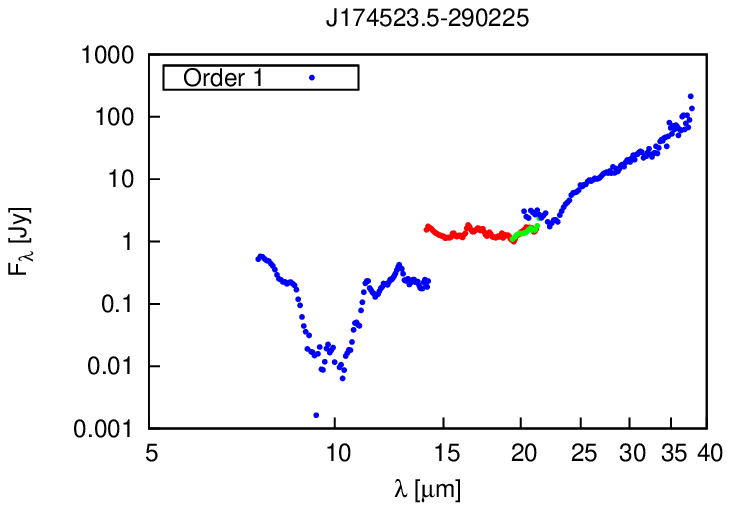}}\\	
	\subfloat{\includegraphics[width=7.5cm]{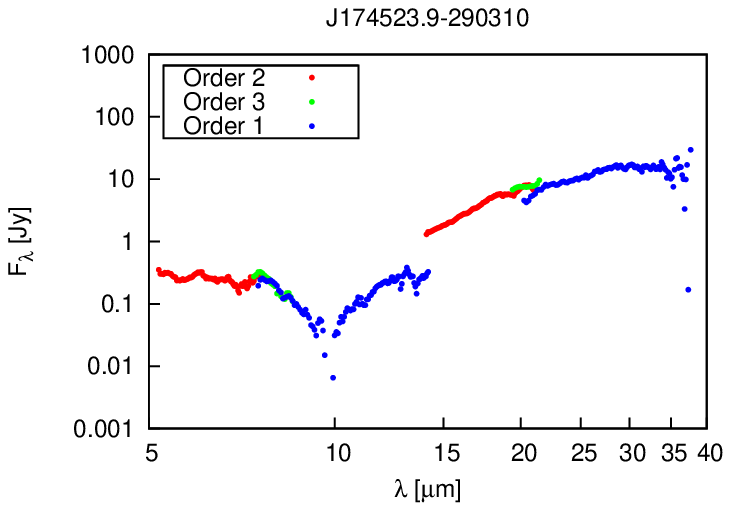}}
	\subfloat{\includegraphics[width=7.5cm]{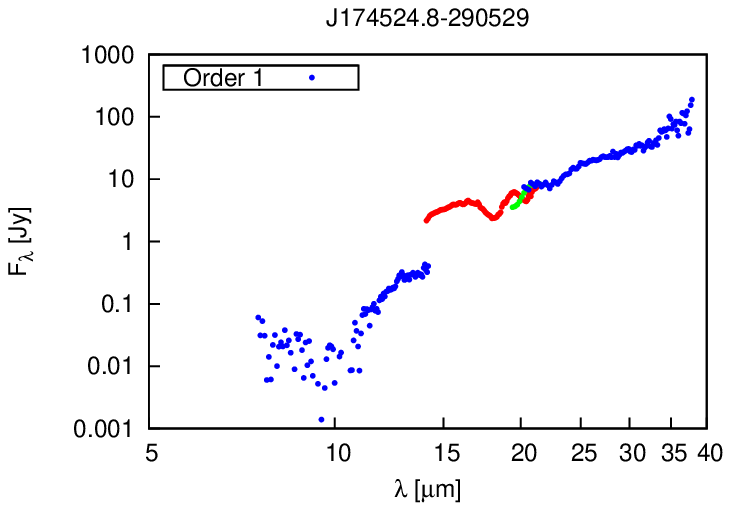}}\\
	\subfloat{\includegraphics[width=7.5cm]{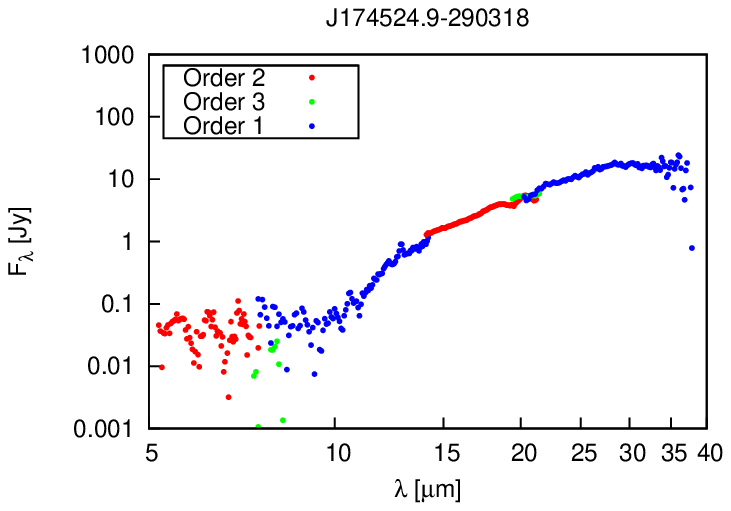}}
	\subfloat{\includegraphics[width=7.5cm]{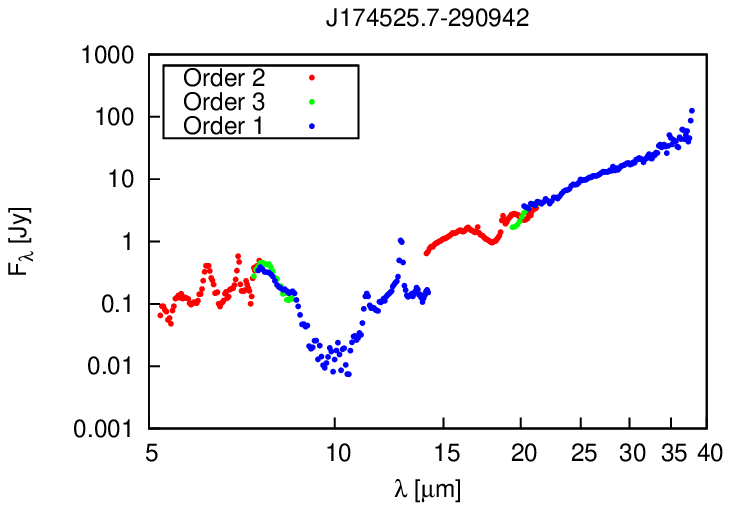}}
\end{figure*}

\addtocounter{figure}{-1}
\begin{figure*}
	\centering		
	\caption{Continued}
	\subfloat{\includegraphics[width=7.5cm]{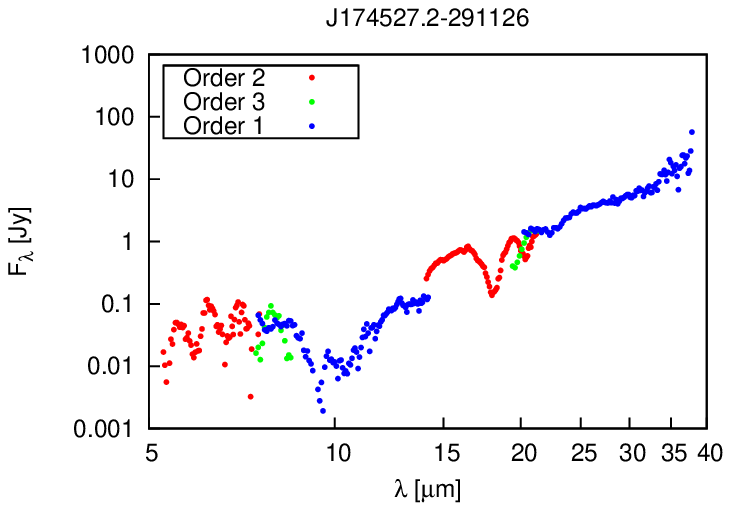}}
	\subfloat{\includegraphics[width=7.5cm]{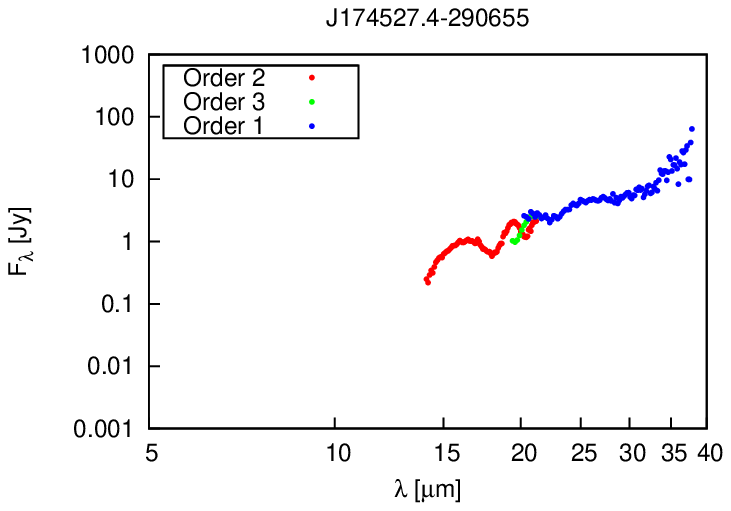}}\\		
	\subfloat{\includegraphics[width=7.5cm]{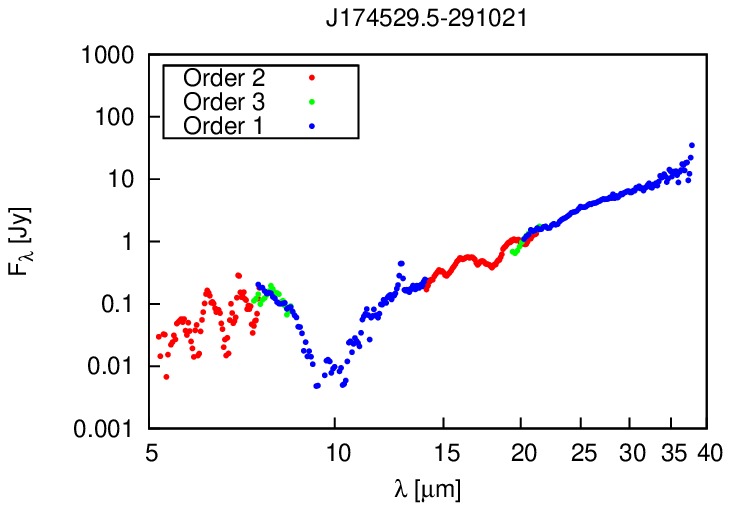}}
	\subfloat{\includegraphics[width=7.5cm]{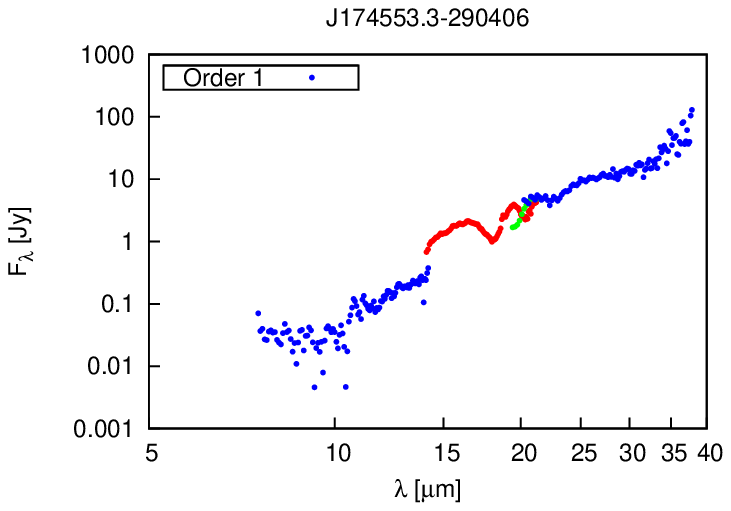}}\\
	\subfloat{\includegraphics[width=7.5cm]{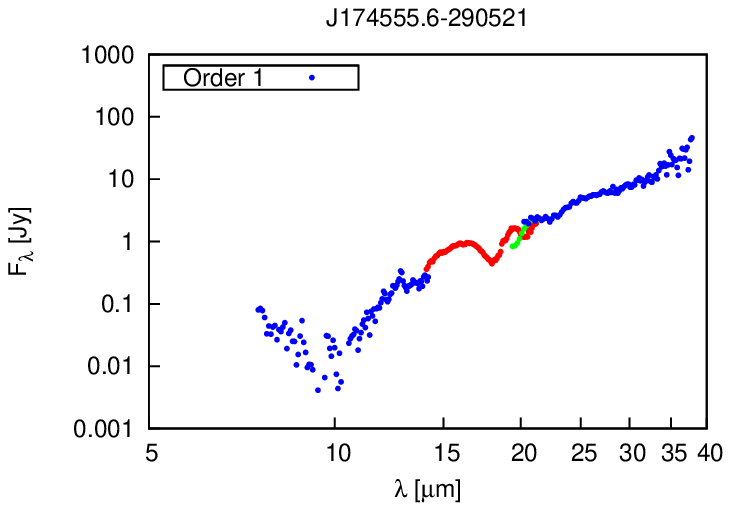}}	
	\subfloat{\includegraphics[width=7.5cm]{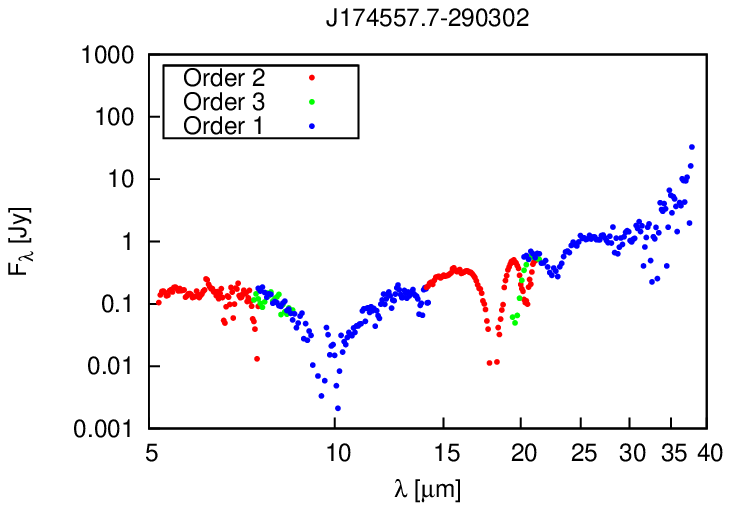}}\\
	\subfloat{\includegraphics[width=7.5cm]{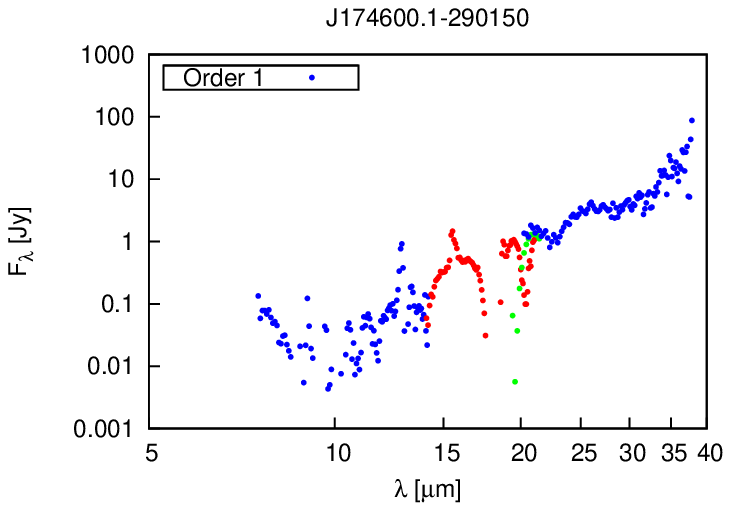}}
	\subfloat{\includegraphics[width=7.5cm]{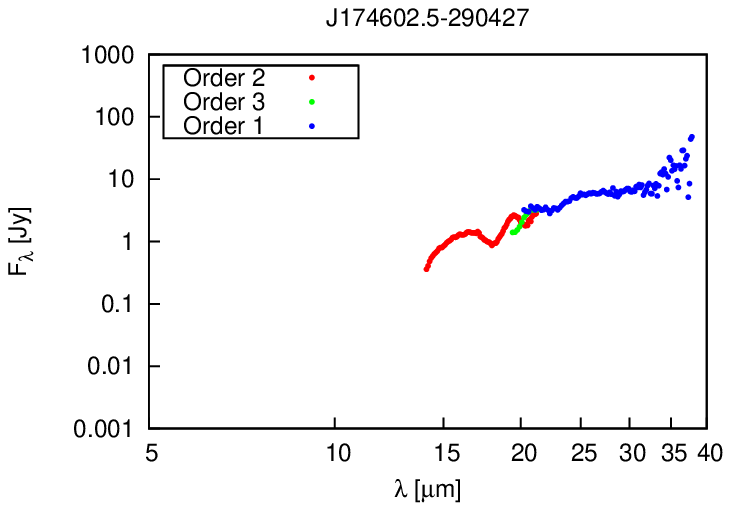}}
\end{figure*}

\addtocounter{figure}{-1}
\begin{figure*}
	\centering			
	\caption{Continued}
	\subfloat{\includegraphics[width=7.5cm]{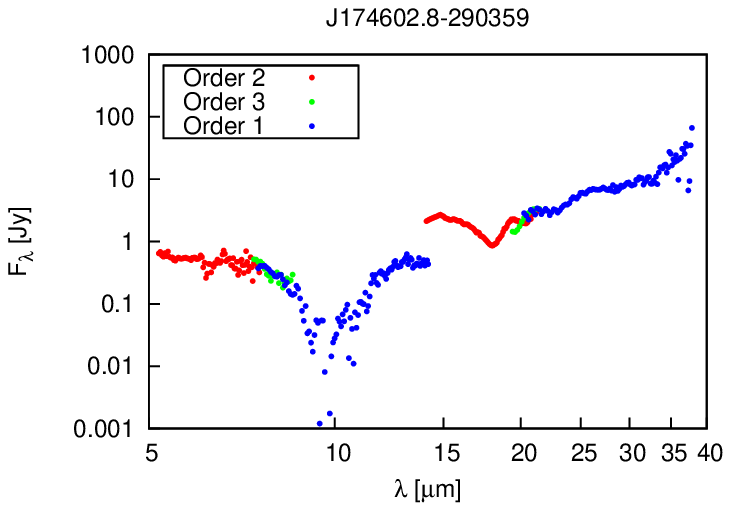}}
	\subfloat{\includegraphics[width=7.5cm]{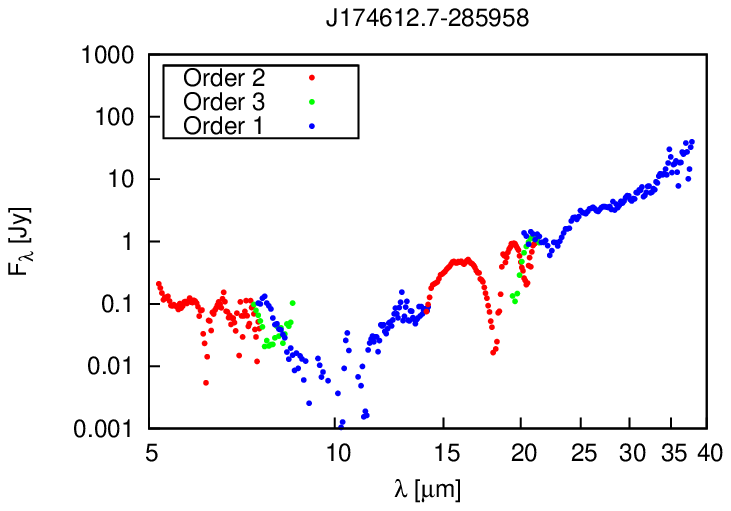}}\\
	\subfloat{\includegraphics[width=7.5cm]{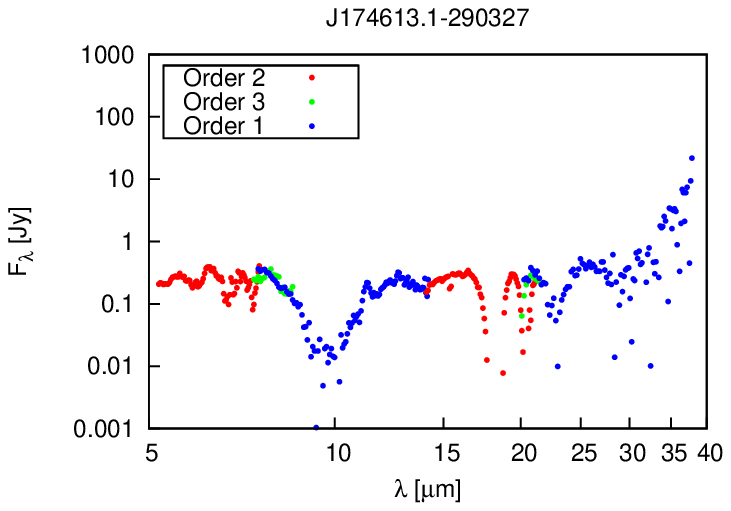}}
	\subfloat{\includegraphics[width=7.5cm]{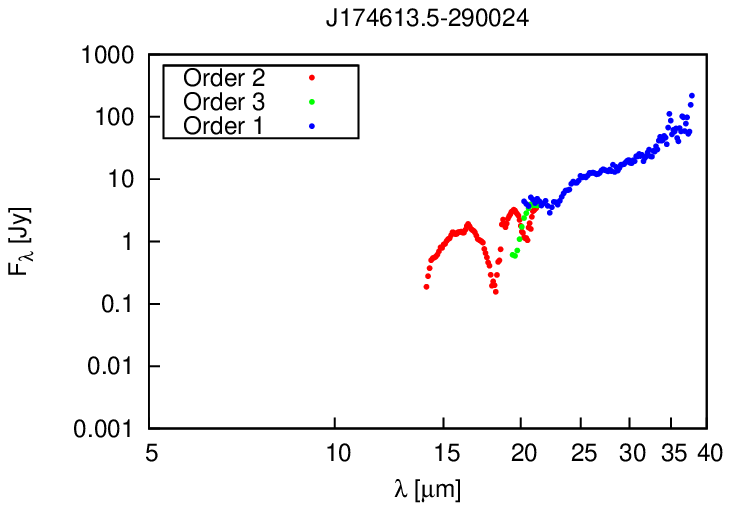}}\\
	\subfloat{\includegraphics[width=7.5cm]{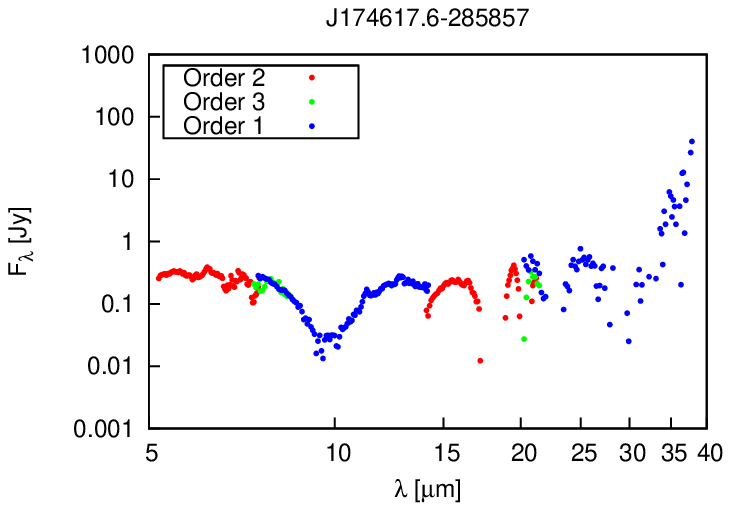}}
	\subfloat{\includegraphics[width=7.5cm]{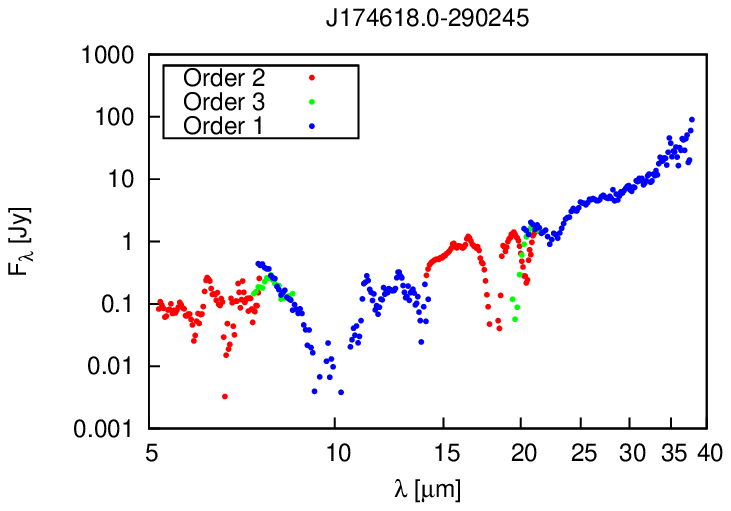}}\\
	\subfloat{\includegraphics[width=7.5cm]{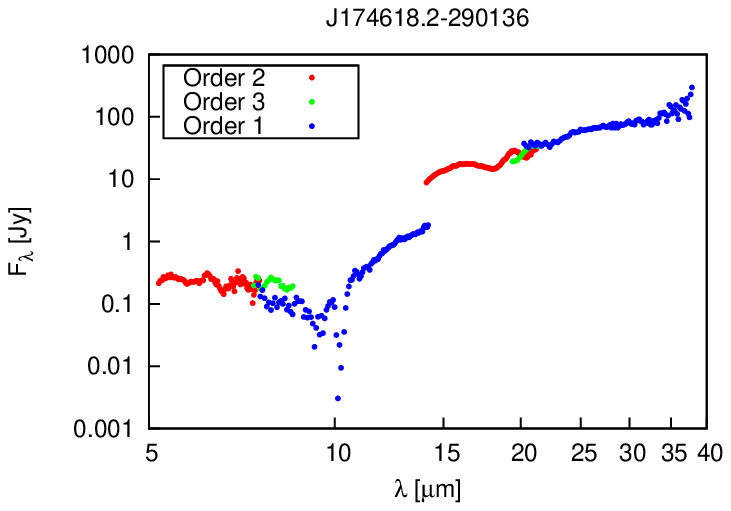}}			
	\subfloat{\includegraphics[width=7.5cm]{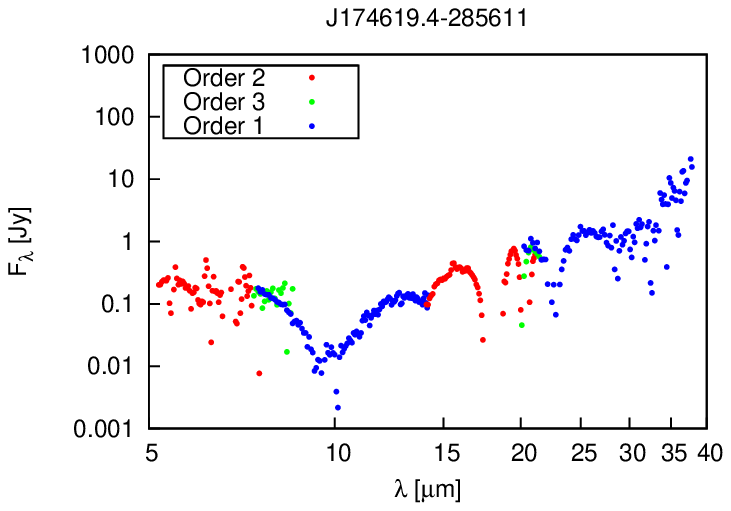}}
\end{figure*}

\addtocounter{figure}{-1}
\begin{figure*}
	\centering
	\caption{Continued}
	\subfloat{\includegraphics[width=7.5cm]{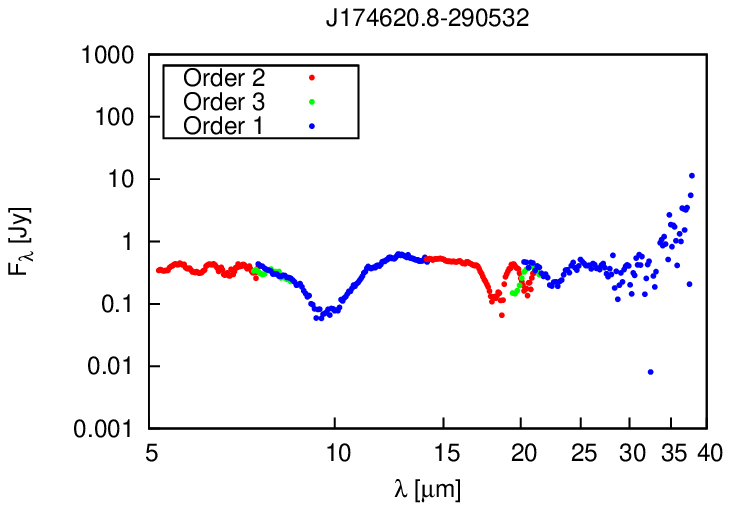}}
	\subfloat{\includegraphics[width=7.5cm]{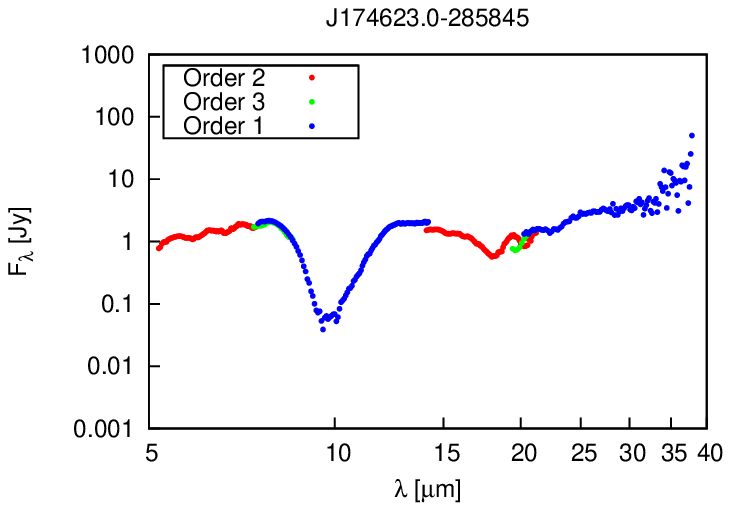}}\\
	\subfloat{\includegraphics[width=7.5cm]{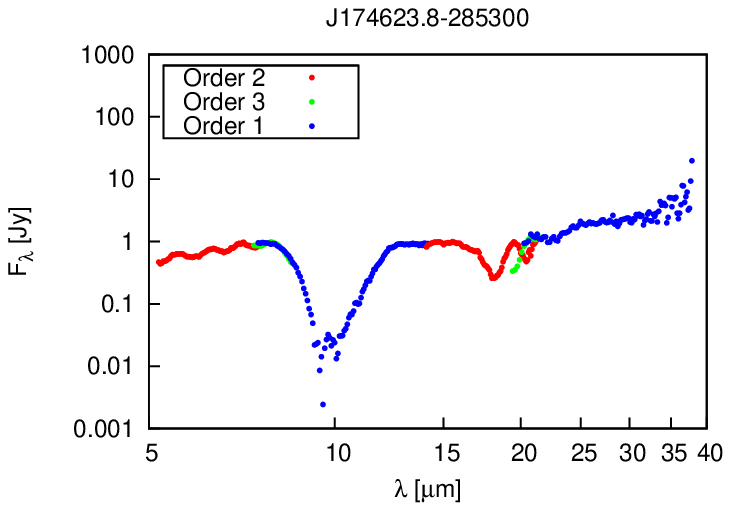}}		
	\subfloat{\includegraphics[width=7.5cm]{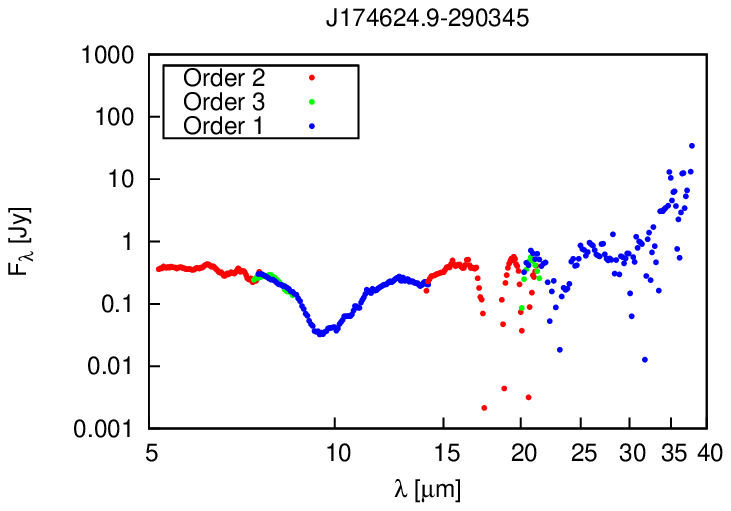}}\\
	\subfloat{\includegraphics[width=7.5cm]{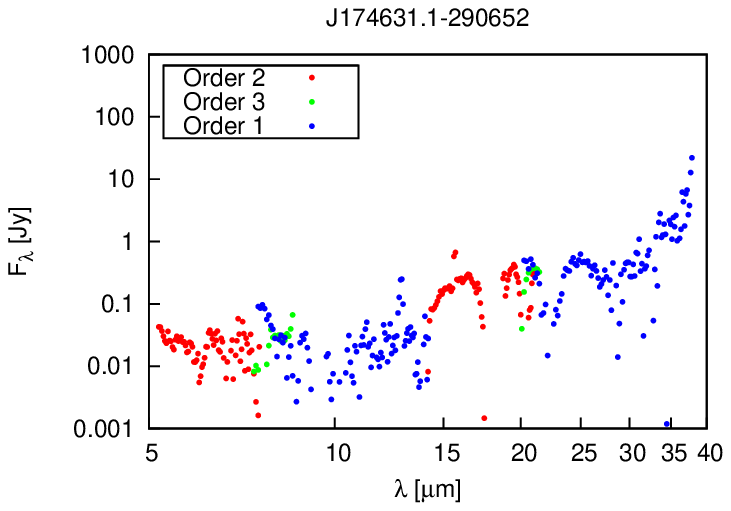}}	
	\subfloat{\includegraphics[width=7.5cm]{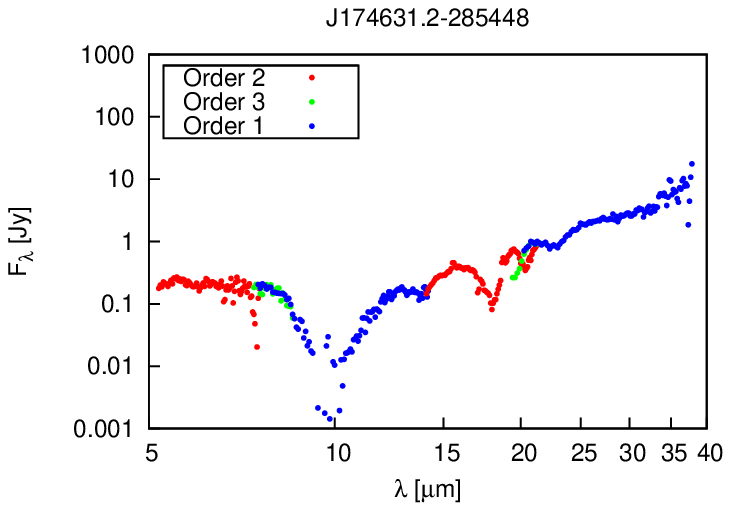}}\\
	\subfloat{\includegraphics[width=7.5cm]{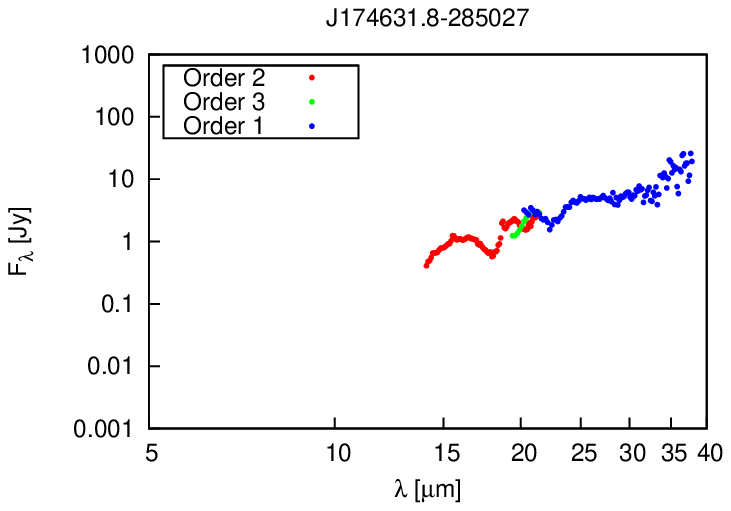}}
	\subfloat{\includegraphics[width=7.5cm]{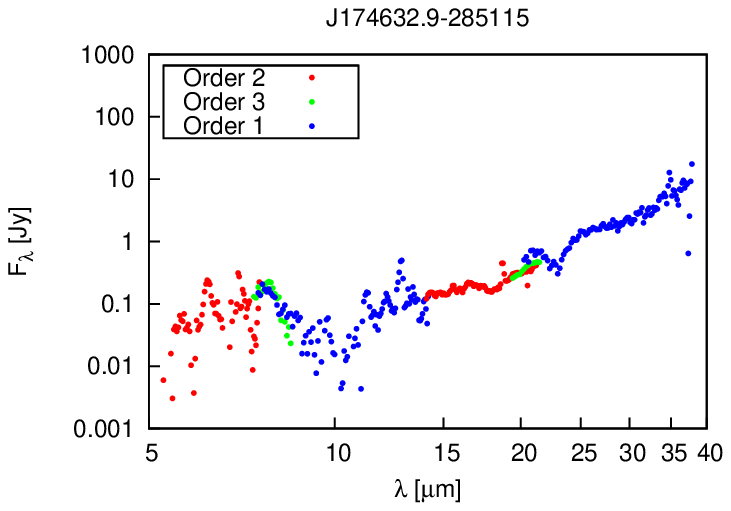}}	
\end{figure*}			
}

\subsubsection{Classification of the 57 sources}

The vast majority of ISOGAL sources brighter than 5~mag at 15~$\mu$m are either of young stellar nature or late-type evolved stars (Paper I). While young sources are characterized by very steep rising continua in the infrared range, late-type evolved stars show more flat to falling continua. An inspection of our spectra shows that some of the sources seem to be composed of a falling SL and a rising LL spectrum.

The two sources that were identified as OH/IR stars by \citep{Sjouwerman1998} both show flat spectra. The only detectable spectral features are the silicate absorption bands at 9.8 and 18~$\mu$m. Neither PAH emission bands nor forbidden fine structure lines are visible in the spectra. 

In order to identify the sources as young or late-type evolved objects, we looked at the slope of the spectrum as well as the detected spectral features. To allow for a quantitative classification of the spectral slopes, we determined the ratio of the flux densities at 33 and 5.3~$\mu$m, $\frac{F_{33~\mu \rm m}}{F_{5.3~\mu \rm m}}$. The fluxes at these two wavelengths were selected in order to obtain the slope over a large part of the spectrum (both SL and LL parts) and to avoid spectral features at the short wavelengths end and the large flux jumps between data points at the long wavelength end. We expect large values for this parameter for young sources with a steep rising spectrum and smaller values for more evolved sources with a flat spectrum. The spectra of sources that showed a flux jump between the SL and the LL module were first corrected for the mismatch and then the slope parameter was determined for these sources. The two OH/IR stars in our test sample have slope parameter values of 2 and 3. After an inspection of the data, we set the threshold for the identification of a source as a young object to be F$_{33~\mu}$/F$_{5.3~\mu}$~$\geq$~10. Sources that were only detected with the SL1 and the LL modules or only the LL modules were identified by comparing their spectra to the other sources.

In dependence of the slope parameter, we classified the sources as young objects or late-type evolved objects. If possible, we subdivided the group of young objects in young stellar objects or HII regions in dependence of the detection of forbidden fine structure lines. In the spectra of young stellar objects or late-type evolved stars we do not expect the detection of PAH emission bands or forbidden fine structure lines. In fourteen of the sources, the identification of the spectral features is not clear. However, the slope of their spectra classifies them as young objects.

\begin{itemize}
 	\item Young object (YO) (49 sources)
 		\begin{itemize}
			\item Young stellar object (YSO) (14 sources)
			\begin{itemize}
				\item Steep rising spectrum
				\item No PAH features or forbidden fine structure lines
			\end{itemize}
			\item \ion{H}{II} region (\ion{H}{II}) (21 sources)
			\begin{itemize}
				\item Steep rising spectrum
				\item Forbidden fine structure lines
			\end{itemize}	
			\item  Steep rising spectrum but no other clear features (14 sources)
 		\end{itemize}
 	\item Late-type evolved object (LEO) (8 sources)
	\begin{itemize}
		\item Flat spectrum
		\item No PAH features or forbidden fine structure lines
	\end{itemize}
\end{itemize}

Furthermore, it should be mentioned that the classification in YSOs and \ion{H}{II} regions is not completely reliable. On one hand, a young \ion{H}{II} region does not have to show emission of forbidden fine structure lines yet. On the other hand, a young stellar object could have forbidden fine structure emission lines in its spectrum, originating from the ambient medium or a nearby source. However, the slope of the spectrum unambiguously identifies these sources as young objects.

Following Paper I, the flux ratio $F_{\rm E}/F_{\rm D}$ of an MSX point source in the E (21~$\mu$m) and D (15~$\mu$m) bands can be used as an identification criterion for young massive stars, if $F_{\rm E}/F_{\rm D}~>~2$. For 21 of our ISOGAL sources, we could find an MSX point source within 8$\arcsec$ (Only MSX sources with flux quality flags of 2--4 were considered). Based on their spectra, eighteen of them were identified as YSO or \ion{H}{II} regions and their MSX counterparts showed $F_{\rm E}/F_{\rm D}$ flux ratios above 2. The three other sources were classified as late-type evolved objects, supported by the flux ratios of their MSX counterparts $F_{\rm E}/F_{\rm D}$ of 0.85, 1.31, and 1.57.

\small
\begin{longtable}{cccccccc}
\caption{\label{Ident} ISOGAL colors and identification of the sources. The first three columns give the source names and their [7] and [15] ISOGAL magnitudes. Column 4 through 6 list the [7]-[15] ISOGAL color, the spatial extent parameter $\sigma_{15}$ and the slope parameter $\frac{F_{33~\mu \rm m}}{F_{5.3~\mu \rm m}}$. In the last two columns, the sources are identified and the detection of radio emission at their position is indicated.}\\ \hline                      
Source name	&	[7]	&	[15]	&	[7]-[15]	&	$\sigma_{15}$	&	F$_{33~\mu m}$/F$_{5.3~\mu m}$	&	Identification	&	Radio source?	\\
                           &      (mag)  & (mag) & & (mag) & & & \\\hline
\endfirsthead 
\caption{continued.}\\	\hline	 
Source name	&	[7]	&	[15]	&	[7]-[15]	&	$\sigma_{15}$	&	F$_{33~\mu m}$/F$_{5.3~\mu m}$	&	Identification	&	Radio source?	\\
                           &      (mag)  & (mag) & & (mag) & & & \\\hline\endhead \hline
\endfoot 															
J174457.4--291003	&	8	&	4.95	&	3.05	&	0.1	&		&	YSO	&	---	\\
J174459.1--290653	&	6.89	&	3.49	&	3.4	&	0.05	&	138	&	YSO	&	---	\\
J174500.7--291007	&	8	&	4.92	&	3.08	&	0.03	&	250	&	YO	&	---	\\
J174501.0--285622	&	8	&	4.83	&	3.17	&	0.19	&		&	\ion{H}{II}	&	---	\\
J174504.9--291146	&	4.27	&	1.97	&	2.3	&	0.11	&	58	&	\ion{H}{II}	&	\checkmark	\\
J174505.1--290937	&	8	&	3.83	&	4.17	&	0.11	&	48	&	YO	&	---	\\
J174505.6--291018	&	5.24	&	1.78	&	3.46	&	0.09	&	678	&	\ion{H}{II}	&	\checkmark	\\
J174506.5--291118	&	5.29	&	0.88	&	4.41	&	0.07	&	782	&	\ion{H}{II}	&	\checkmark	\\
J174508.0--290655	&	8	&	4.34	&	3.66	&	0.14	&		&	YSO	&	---	\\
J174508.0--291039	&	6.59	&	4.24	&	2.35	&	0.11	&	63	&	\ion{H}{II}	&	---	\\
J174508.1--290840	&	7.69	&	5.11	&	2.58	&	0.06	&	240	&	\ion{H}{II}	&	\checkmark	\\
J174508.7--290348	&	8	&	4.17	&	3.83	&	0.12	&	230	&	\ion{H}{II}	&	---	\\
J174511.3--290621	&	6.44	&	4.31	&	2.13	&	0.07	&	72	&	YO	&	---	\\
J174512.6--290646	&	8	&	4.97	&	3.03	&	0.03	&	70	&	YO	&	---	\\
J174515.4--291213	&	8	&	3.4	&	4.6	&	0.11	&	1242	&	YSO	&	---	\\
J174515.9--290155	&	8	&	2.63	&	5.37	&	0.13	&	246	&	YSO	&	---	\\
J174516.2--290315	&	5.66	&	1.03	&	4.63	&	0.19	&	163	&	YSO	& ---	\\
J174517.1--291155	&	8	&	5.02	&	2.98	&	0.09	&		&	YSO	&	---	\\
J174517.8--290813	&	7.3	&	4.1	&	3.2	&	0.09	&	290	&	YO	&	---	\\
J174518.1--290439	&	6.3	&	2.54	&	3.76	&	0.05	&	139	&	\ion{H}{II}	&	\checkmark	\\
J174518.1--291051	&	6.51	&	3	&	3.51	&	0.08	&	2285	&	\ion{H}{II}	&	\checkmark	\\
J174518.3--290628    &       8       &	5.18  &	2.82  &	0.1 	& 	22 	& 	YSO 	& ---\\
J174520.0--290750	&	7.34	&	5.09	&	2.25	&	0.08	&		&	YSO	&	---	\\
J174520.1--290638	&	8	&	4.93	&	3.07	&	0.12	&		&	YO	&	---	\\
J174520.7--290213	&	4.57	&	0.4	&	4.17	&	0.12	&	554	&	YO	&	---	\\
J174522.1--291059	&	6.81	&	4.61	&	2.2	&	0.04	&	124	&	\ion{H}{II}	&	---	\\
J174522.4--290242    & 	8	& 	3.68 	& 	4.32	& 	0.17 	& 	109 	& \ion{H}{II} & ---\\
J174523.3--290331	&	5.52	&	2.29	&	3.23	&	0.14	&	24	&	YO	&	---	\\
J174523.5--290225	&	8	&	4.62	&	3.38	&	0.14	&		&	YO	&	---	\\
J174523.9--290310	&	6.09	&	3.29	&	2.8	&	0.14	&	10	&	YO	&	---	\\
J174524.8--290529	&	8	&	3.89	&	4.11	&	0.19	&		&	YSO	&	---	\\
J174524.9--290318	&	8	&	3.33	&	4.67	&	0.14	&	386	&	\ion{H}{II}	&	---	\\
J174525.7--290942	&	8	&	4.68	&	3.32	&	0.15	&	73	&	\ion{H}{II}	&	---	\\
J174527.2--291126	&	7.54	&	5.15	&	2.39	&	0.1	&	73	&	YSO	&	---	\\
J174527.4--290655	&	8	&	5.12	&	2.88	&	0.13	&		&	YSO	&	---	\\
J174529.5--291021	&	6.89	&	4.51	&	2.38	&	0.07	&	271	&	\ion{H}{II}	&	---	\\
J174553.3--290406	&	8	&	3.55	&	4.45	&	0.09	&		&	YO	&	---	\\
J174555.6--290521	&	7.05	&	3.77	&	3.28	&	0.09	&		&	\ion{H}{II}	&	---	\\
J174557.7--290302	&	7.53	&	5.05	&	2.48	&	0.06	&	11	&	YO	&	---	\\
J174600.1--290150	&	8	&	3.48	&	4.52	&	0.06	&		&	\ion{H}{II}	&	---	\\
J174602.5--290427	&	8	&	3.81	&	4.19	&	0.19	&		&	YSO	&	---	\\
J174602.8--290359	&	5.9	&	2.43	&	3.47	&	0.08	&	2	&	LEO	&	---	\\
J174612.7--285958	&	7.42	&	5.08	&	2.34	&	0.14	&	71	&	\ion{H}{II}	&	---	\\
J174613.1--290327	&	8	&	4.75	&	3.25	&	0.13	&	2	&	LEO	&	---	\\
J174613.5--290024	&	6.84	&	4.52	&	2.32	&	0.22	&		&	\ion{H}{II}	&	---	\\
J174617.6--285857 	& 	8	&	4.81 	& 	3.19 	&	0.13 	& 	1 	& 	LEO 	& --- \\
J174618.0--290245	&	8	&	5.24	&	2.76	&	0.21	&	186	&	\ion{H}{II}	&	---	\\
J174618.2--290136	&	6.35	&	0.88	&	5.47	&	0.19	&	62	&	YSO	&	---	\\
J174619.4--285611 	& 	6.98 	& 	4.57 	& 	2.41 	& 	0.19 	& 	6 	& 	LEO 	& ---\\
J174620.8--290532	&	6.13	&	4.22	&	1.91	&	0.04	&	1	&	LEO	&	---	\\
J174623.0--285845	&	5.07	&	2.97	&	2.1	&	0.05	&	3	&	LEO	&	---	\\
J174623.8--285300	&	5.21	&	3.2	&	2.01	&	0.03	&	6	&	LEO	&	---	\\
J174624.9--290345	&	8	&	5.23	&	2.77	&	0.09	&	1	&	LEO	&	---	\\
J174631.1--290652	&	8	&	5.14	&	2.86	&	0.09	&	47	&	YO	&	---	\\
J174631.2--285448	&	6.9	&	4.76	&	2.14	&	0.13	&	17	&	YO	&	---	\\
J174631.8--285027	&	8	&	4.12	&	3.88	&	0.21	&		&	\ion{H}{II}	&	--- \\
J174632.9--285115	&	7.67	&	5.21	&	2.46	&	0.14	&	219	&	\ion{H}{II}	&	---	\\
\end{longtable} 

\twocolumn
\normalsize

In addition, we looked for radio emission in the vicinity of our ISOGAL sources in images of the VLA archive at 1.4, 4.7, and 8.4~GHz in order to identify potential \ion{H}{II} regions. At the positions of six ISOGAL sources, which were classified as \ion{H}{II} regions on the basis of their spectra, radio emission was detected. Several other sources are located at the edge of diffuse radio emission. 

Table \ref{Ident} shows the ISOGAL colors as well as the spectral identification of the sources. Columns 1--5 list the sources names together with their ISOGAL magnitudes at 7 and 15 $\mu$m, the [7]--[15] color and the spatial extent parameter $\sigma_{15}$. The sixth column contains the slope parameter $\frac{F_{33~\mu \rm m}}{F_{5.3~\mu \rm m}}$. In the last two columns the identification of the sources as well as the detection of radio emission at the positions of the sources in the VLA images are reported.

\subsection{Test of the ISOGAL selection criteria}

\label{DiscussionPartI}

One of the intentions of this publication was to determine if the applied photometric color criteria ([15]~<~5.25, [7]--[15]~>~1.8) permit the selection of only young objects in a sample of ISOGAL point sources. The identification of eight late-type evolved objects in our test sample shows that the color criteria do not constrain the group of sources to objects of young stellar nature only. 

We confirmed that a combination of the spatial extension of the sources at 15~$\mu$m, characterized by the parameter $\sigma_{15}$, with their [7]--[15] colors gives a reliable method to distinguish between young and late-type evolved objects.

\begin{figure}
	\centering
		\includegraphics[width=9cm]{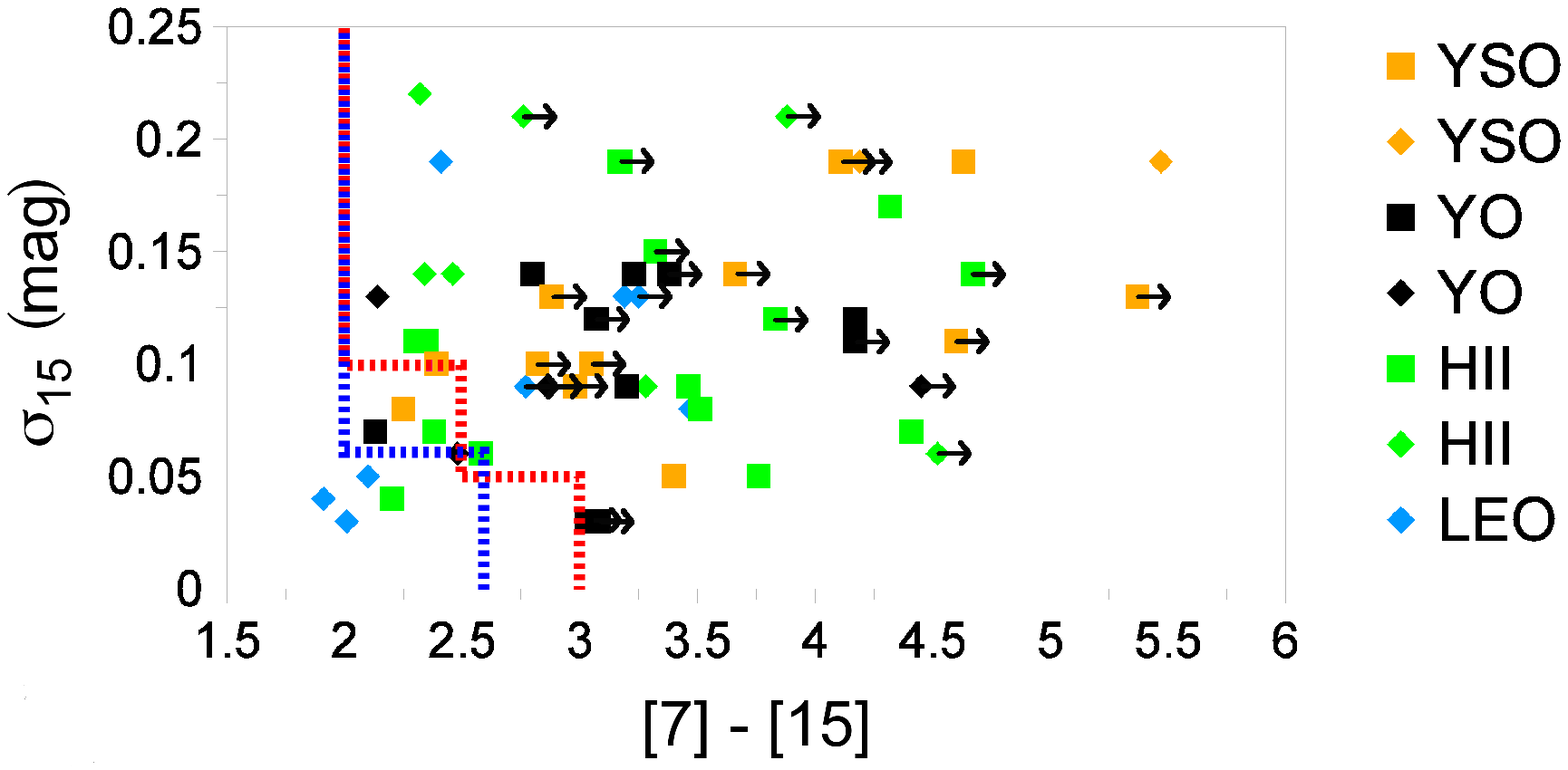}
	\caption{The image shows the spatial extension parameter $\sigma_{15}$ plotted against the [7]-[15] color. The arrows indicate lower limits of the [7]-[15] colors for the sources which were only detected at 15~$\mu$m. The squares mark the sources whose [7] and [15] colors were not observed at the same time, diamonds highlight sources which were simultaneously observed at 7 and 15~$\mu$m. The red line represents the selection criteria of Paper I, the blue line the selection criteria which were defined in this publication. The image shows that color selection criteria which are combined with the spatial extension parameter $\sigma_{15}$ permit the distinction between young and evolved objects with a low contamination rate of late-type evolved objects.}
	\label{sigma15vs7-15}
\end{figure}

In Figure \ref{sigma15vs7-15}, $\sigma_{15}$ is plotted against the [7]--[15] colors. Sources which were only detected at 15~$\mu$m are marked with an arrow. The image shows that young objects and late-type evolved stars cannot be divided into two separate sections in the plot. However, criteria which are based on the ISOGAL colors as well as the spatial extension parameter allow the extraction of young objects from a mixed sample with a small contamination by evolved sources. The red line represents the selection criteria 
	\[2 \leq [7] - [15] < 2.5 \ \textnormal{and} \ \sigma_{15} > 0.1~\textnormal{mag}\]
	\[2.5 \leq [7] - [15] < 3 \ \textnormal{and} \ \sigma_{15} > 0.05~\textnormal{mag}\]
	\[[7] - [15] \geq 3, \textnormal{no constraint on} \ \sigma_{15}\]

which were defined in Paper I from a group of ISOGAL sources with known evolutionary states. Applying these criteria to our test sample results in the selection of 44 young objects and five late-type evolved stars, corresponding to a LEO contamination rate of 10.2\% (5/49). In order to reduce the LEO contamination rate by selecting more young objects and to simplify the selection criteria, we refined the criteria (blue line) to
\[2 \leq [7] - [15] < 2.6 \ \textnormal{and} \ \sigma_{15} \geq 0.06~\textnormal{mag}\]
\[2.6 \leq [7] - [15], \textnormal{no constraint on} \ \sigma_{15}\]

These criteria would yield the same number of late-type evolved objects but 48 young objects, resulting in a LEO contamination rate of 9.4\% (5/53). These criteria also agree with the ISOGAL source sample of Paper I and will be used in the second part of this publication in order to identify young sources in the CMZ in the ISOGAL point source catalogue.

In order to estimate the final LEO contamination rate, also the remaining 11 sources for which no spectrum was assembled and the 27 sources that were identified beforehand as late evolved stars have to be considered. If we apply the defined selection criteria on the 27 rejected sources, nine of them are classified as young objects, increasing the contamination rate to 22.6\% (14/62). All 11 non-identified sources of our test sample of 68 sources are sorted in the group of young objects. The sources labeled as group C sources are probably false detections in the ISOGAL catalogue, increasing the contamination rate to 27.3\% (18/66). Only one of the group D objects has an MSX counterpart within 8$\arcsec$ which has a flux density ratio of $F_{\rm E}/F_{\rm D}$~>~2. The ISOGAL sources in group D are resolved in more than one source at short wavelengths by the Spitzer observations. The ISOGAL colors of these sources can be dominated by either young sources or more evolved objects in the cluster. The contamination rate is therefore 34.2\% (25/73) if the group D objects are defined as more evolved objects and contribute to the contamination rate or 24.7\% (18/73) if they are identified as young objects. In conclusion, the contamination rate with non-young objects is in the range 25 to 35\%. 

Another spectroscopic study of YSO candidates in the CMZ was recently conducted by \citet{An2011}. Based on IRAC colors, they selected 107 point sources for their sample and obtained high-resolution Spitzer/IRS spectra for these targets. The shape of the absorption profile of CO$_2$ ice at 15~$\mu$m served as their identification criterion. In total, they identified 35 YSOs and possible YSOs. 

Since we only have low-resolution data for our sources, we could not test the An et al. identification criterion on our test sample. A comparison of the An et al. source sample and our test sample showed that both have three sources in common. An et al. identified only one of them as a possible YSO (source J174529.5-291021) on the basis of the CO$_{2}$ ice absorption profile. Beside this ice absorption band, our Spitzer/IRS spectrum of this source also shows emission lines of [ArII] and [NeII] in its spectrum. Thus, we identified this source as an \ion{H}{II} region. We conclude that this source could be a transition object between YSOs and \ion{H}{II} regions, showing signatures of both phases. We also identified the other two sources as \ion{H}{II} regions. An et al. exclude that these sources are YSOs.

While the An et al. study focuses only on YSOs, we will obtain a broader picture of the star formation history in the CMZ since we also include objects in our test sample that are in later evolutionary stages of the star formation process, such as \ion{H}{II} regions.

\section{Star formation in the Central Molecular Zone}

In this part, we will apply the defined selection criteria on all ISOGAL sources that have been detected in the central molecular zone in order to determine the star formation activity in the central 450~pc of our Galaxy. Young object candidates are selected in an area covering $\pm$~1.5$\degr~\times~\pm$~0.5$\degr$ around the Galactic Center. The list of ISOGAL sources in the CMZ will be compared with the MSX point source catalogue (PSC) and MSX sources with a flux density ratio of $F_{\rm E}/F_{\rm D}$~$\geq$~2 without an ISOGAL counterpart within 8$\arcsec$ will be added to the CMZ source list of young object candidates.

We will derive bolometric luminosities for all sources in the CMZ sample from their flux densities at 15~$\mu$m by using $\frac{L_{\rm bol}}{F_{15}}$  conversion factors which will be obtained from our test sample sources. Furthermore, we will determine mass estimates for all CMZ sample sources in order to determine the total mass in young objects that has been formed over the last $\sim$1~Myr and thus derive an average star formation rate for the CMZ.

\subsection{Selection of the Candidate Young Objects}

\subsubsection{ISOGAL point sources}

\begin{table}[h]
\caption{\label{Ifields}ISOGAL fields used in the present analysis. The fields names (Col. 1) give the galactic coordinates of their centers. Column 2 and 3 contain the LW filter numbers at 7 and 15~$\mu$m. The last two columns show the field of view (FOV) and the pixel size of the field.}
\begin{center}
\begin{tabular}{lllll}
Name & \multicolumn{2}{c}{Filters} & FOV & Pixel\\\hline
FC--00121--00003	& LW5 & LW9 & 9${\arcmin}$ x 17${\arcmin}$ 	& 6$\arcsec$ \\
FC--00112--00035	& LW6 & LW9 & 28${\arcmin}$ x 20${\arcmin}$	& 6$\arcsec$ \\
FC--00109+00031	& LW6 & LW9 & 34${\arcmin}$ x 23${\arcmin}$ 	& 6$\arcsec$ \\
FC--00090--00003 	& LW5 & LW9 & 21${\arcmin}$ x 17${\arcmin}$	& 6$\arcsec$ \\
FC--00062--00006 	& LW5 & LW9 & 12${\arcmin}$ x 17${\arcmin}$	& 3$\arcsec$ \\
FC--00062--00040	& LW6 & LW9 & 15${\arcmin}$ x 12${\arcmin}$	& 6$\arcsec$ \\
FC--00039+00018  	& LW6 & LW9 & 39${\arcmin}$ x 9${\arcmin}$		& 6$\arcsec$ \\
FC--00027--00006 	& LW5 & LW9 & 19${\arcmin}$ x 17${\arcmin}$	& 3$\arcsec$ \\
FC+00004+00040   	& LW6 & LW9 & 46${\arcmin}$ x 14${\arcmin}$	& 6$\arcsec$ \\
FC+00005--00024  	& LW5 & LW9 & 18${\arcmin}$ x 15${\arcmin}$	& 6$\arcsec$ \\
FC+00034--00005  	& LW5 & LW9 & 14${\arcmin}$ x 16${\arcmin}$	& 3$\arcsec$ \\
FC+00037+00017   	& LW5 & LW9 & 34${\arcmin}$ x 10${\arcmin}$	& 6$\arcsec$ \\
FC+00059+00002   	& LW5 & LW9 & 14${\arcmin}$ x 7${\arcmin}$		& 3$\arcsec$ \\
FC+00062--00014  	& LW5 & LW9 & 13${\arcmin}$ x 6${\arcmin}$		& 3$\arcsec$ \\
FC+00066--00041  	& LW6 & LW9 & 46${\arcmin}$ x 12${\arcmin}$	& 6$\arcsec$ \\
FC+00067+00038   	& LW6 & LW9 & 19${\arcmin}$ x 14${\arcmin}$	& 6$\arcsec$ \\
FC+00089--00009  	& LW6 & LW9 & 18${\arcmin}$ x 20${\arcmin}$	& 6$\arcsec$ \\
FC+00124--00032   	& LW6 & LW9 & 23${\arcmin}$ x 22${\arcmin}$	& 6$\arcsec$ \\
FC+00127+00035   	& LW6 & LW9 & 23${\arcmin}$ x 20${\arcmin}$	& 6$\arcsec$ \\
\end{tabular}
\end{center}
\end{table}

\begin{figure}
	\centering
		\includegraphics[width=9cm]{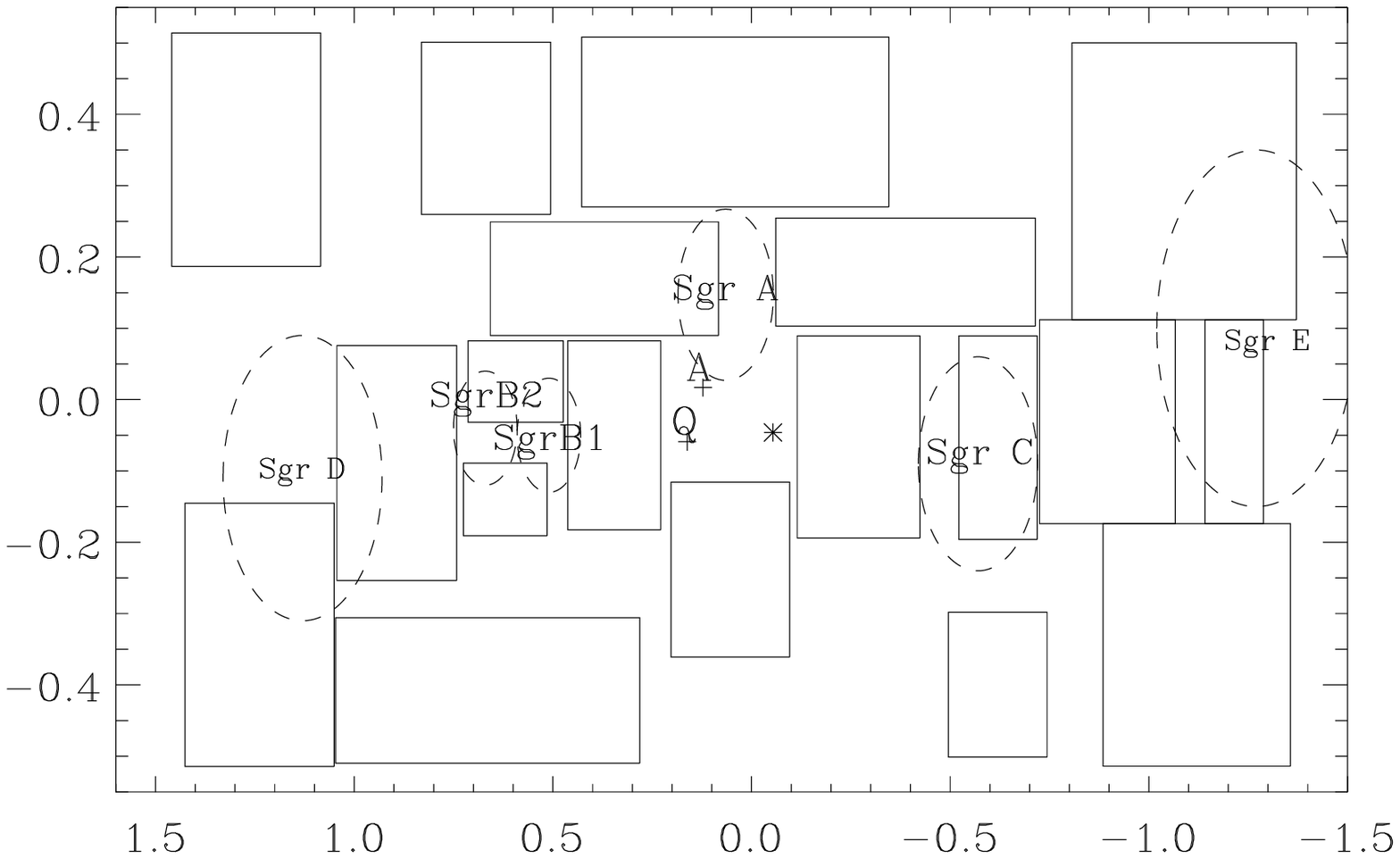}
		\caption{Boundaries of the ISOGAL fields used for the present study (see
also Table \ref{Ifields}), in Galactic coordinates. Some well known regions are indicated: Sgr
A$^*$ (asterisk), the Arches cluster (A), the Quintuplet cluster (Q), as well as the
Sgr A, Sgr B1, Sgr B2, Sgr C, Sgr D and Sgr E star forming regions.}
	\label{IfieldsCMZ}
\end{figure}

The data processing of the ISOGAL survey \citep{Omont2003} and the creation of the ISOGAL--DENIS point source catalogue have been extensively described in \citet{Schuller2003}. Here we will use only ISOGAL fields that have been observed at 7 and 15~$\mu$m. A list of the fields (Col. 1) with which the sources were observed, the filter numbers (Col. 2, 3), field of view of the fields (Col. 4) and the pixel scales of the fields (Col. 5) are given in Table \ref{Ifields}.

The ISOGAL observations were designed to exclude strong IRAS sources, with F$_{12~\mu m}$~$\geq$~6~Jy in general, to avoid saturation effects of the detector. However, for most fields observed around the Galactic Center, narrow filters and 3$\arcsec$ pixels were used, to allow the observations of brighter sources (up to F$_{12~\mu \rm m}$~$\sim$~20 Jy); nevertheless, the Galactic Center itself and a couple of other brightest star forming regions could not be imaged. Fig. \ref{IfieldsCMZ} shows the boundaries of the ISOGAL fields, listed in Table \ref{Ifields}. In addition, known star forming regions are indicated.

As shown in the first part of this publication as well as in Paper I, source samples based only on color selection criteria are contaminated with some post-main sequence stars (planetary nebulae, OH/IR stars with extreme mass losses) although most of the sources are associated with massive stars. Combining a color selection criterion with an indication of spatial extension (i.e., removing the point-like sources) allows to select infrared sources which are very likely to be massive (young) objects. 

The application of the selection criteria defined in the first part of this publication to the ISOGAL fields listed in Table \ref{Ifields} results in the selection of 485 ISOGAL sources with [15]~<~5.25~mag. As described in subsection \ref{SourSel} and \ref{ReducIRS}, a lower limit for the [7] magnitude was assigned to sources that were not detected at 7~$\mu$m. 

\subsubsection{MSX point sources}

The Midcourse Space Experiment performed a survey of the Galactic plane at four mid-infrared wavelengths: 8, 12, 15 and 21~$\mu$m, with a pixel scale of 18$\arcsec$. The sensitivity of the MSX catalogue version 2.3 \citep{Egan2003} is of order 50~mJy at 8~$\mu$m and 1--2~Jy at 15 and 21~$\mu$m.

As has been shown in Paper I, the longest wavelengths of MSX provide another color criterion ($F_{\rm E}/F_{\rm D}$~$\geq$~2) to select probable young massive objects, with little contamination by late-type evolved stars and planetary nebulae \citep[see also][]{Lumsden2002}. In order to have reliable flux density values for the MSX sources at 15 and 21~$\mu$m, we only considered sources with flux quality flags of 2--4 in the D and E bands. Besides the ISOGAL sources, we selected MSX point sources in the CMZ without an ISOGAL counterpart within 8$\arcsec$ that fulfilled the specified MSX color criterion, resulting in 656 MSX sources.

\begin{figure*}
	\centering
		\includegraphics[width=18cm]{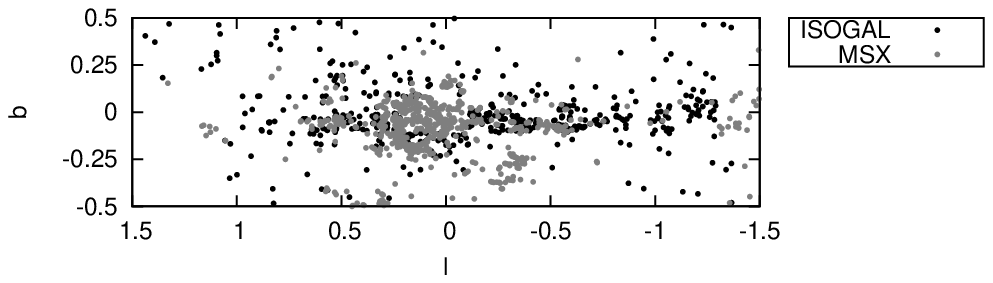}
		\caption{Areal coverage of our CMZ sample with Galactic longitude l and Galactic latitude b. Black dots mark the ISOGAL sources, grey dots the MSX sources without ISOGAL counterparts. The image shows that most of the MSX sources are located at 0$\degr$~$\leq$~l~$\leq$~0.5$\degr$, which could not be observed by ISOGAL because of detector saturation. Most of the sources in the CMZ sample are confined to the Galactic plane (|b|~$\leq$~0.2$\degr$).}
	\label{YOs-CMZ}
\end{figure*}

In total, our CMZ sample contains 1141 young object candidates. Fig. \ref{YOs-CMZ} shows the distribution of the CMZ source sample in Galactic longitude $l$ and Galactic latitude $b$. ISOGAL sources and MSX sources are marked as black and grey dots, respectively. Most of the sources in the CMZ sample are confined to the Galactic plane (|b|~$\leq$~0.2$\degr$) and most of the MSX sources are located at 0$\degr$~$\leq$~l~$\leq$~0.5$\degr$, which could not be observed by ISOGAL because of detector saturation.

\subsection{Discussion}
\label{DiscussionPartII}

\begin{figure*}
\centering
	\subfloat{\includegraphics[width=9cm]{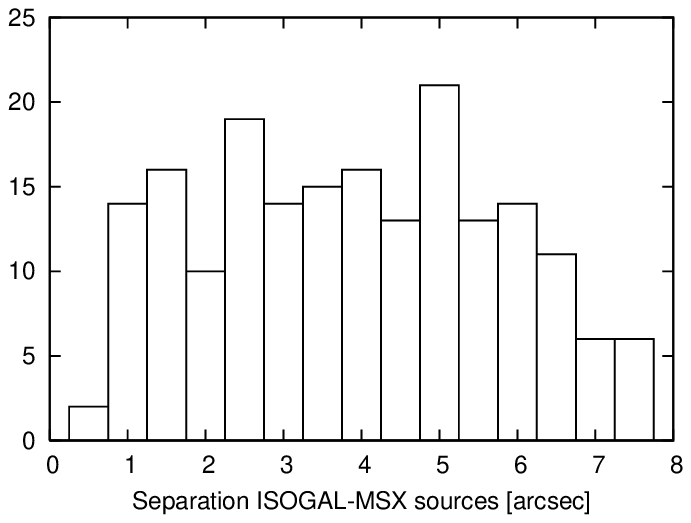}}
	\subfloat{\includegraphics[width=9cm]{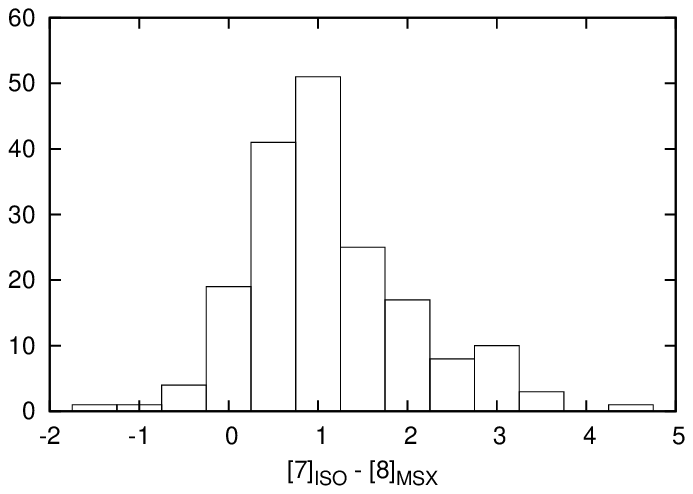}}\\	
	\subfloat{\includegraphics[width=9cm]{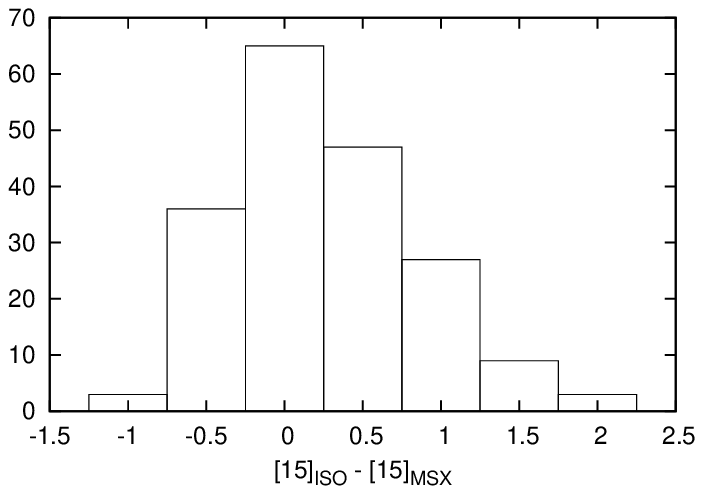}}
	\subfloat{\includegraphics[width=9cm]{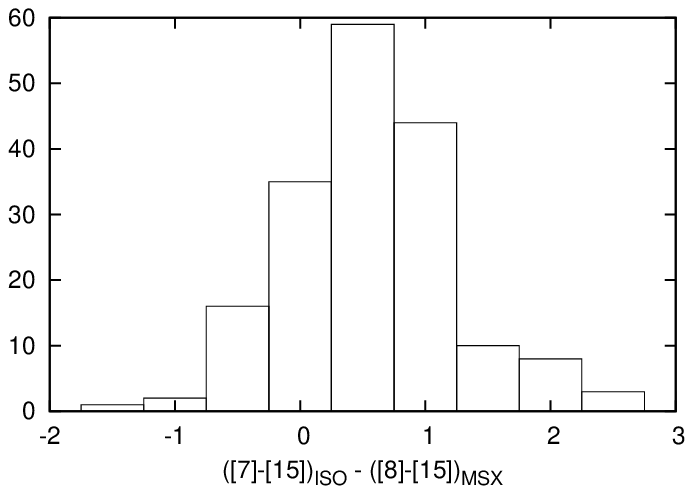}}	
\caption{From top-left to bottom-right, distributions of: Angular separation between the ISOGAL and the associated MSX source, ISO--MSX 7~$\mu$m magnitude differences, ISO--MSX 15~$\mu$m magnitude differences, and ISO--MSX differences in colors.}
\label{ISO-MSX-Colors}
\end{figure*}

Among the 485 ISOGAL sources, 190 (almost 40\%) have an MSX counterpart within a search radius of 8$\arcsec$. The distributions of ISO-MSX position separations (Fig. \ref{ISO-MSX-Colors}) does not show any increase with larger separations, as would be expected from random associations. In addition, the agreement between ISO and MSX magnitudes is generally good. Thus, we do not expect a large contribution by false associations in our CMZ sample. However, the MSX magnitudes tend to be lower (the sources appear brighter) than the ISO sources, especially at 7/8~$\mu$m. This can partly be explained by different central wavelengths (8~$\mu$m ~vs. 7~$\mu$m) that can make a difference for these red objects. Some part of the observed difference may also be due to a Malmquist bias: with a sensitivity more limited for MSX than ISOGAL, the brighter MSX sources will preferentially be associated. In addition, the lower spatial resolution of MSX compared to ISO can also introduce a bias in that direction, when a single MSX source is resolved into several ISOGAL point sources.

\begin{figure}
	\centering
		\includegraphics[width=9cm]{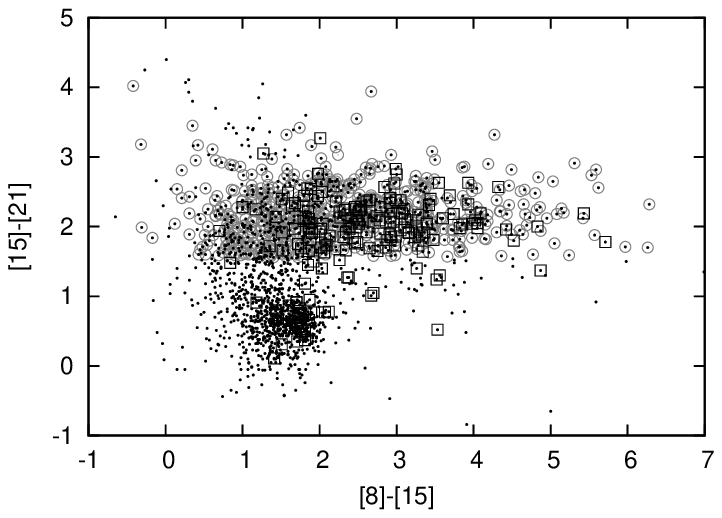}
	\caption{MSX Color-Color diagram. All MSX sources with flux quality flags $\geq$~1 in A and C band and $\geq$~2 in D and E band in the CMZ are plotted with black dots. The black boxes mark all MSX sources with an ISOGAL counterpart within 8$\arcsec$. MSX sources with $F_{\rm E}/F_{\rm D}$~$\geq$~2 without an ISOGAL counterpart within 8$\arcsec$ are shown as grey circles. The image shows that almost half of the MSX sources in the CMZ appear in a `clump', around [8]--[15]~$\sim$~2 and [15]--[21]~$\sim$~0.5, but only $\sim$~20 of the MSX sources with an ISOGAL counterpart are located in this clump, and could actually be `normal' field objects. On the opposite, 80\% of the MSX objects with an ISOGAL counterpart have a $F_{\rm E}/F_{\rm D}$ ratio above 2 and indeed trace a specific population.}
	\label{MSX-Color-Color}
\end{figure}	

Among these 190 MSX sources, 177 have flux quality flags of $\geq$~1 in A and C band and $\geq$~2 in D and E band. Their positions in the [15]--[21] vs. [8]--[15] color-color diagram are shown in Fig. \ref{MSX-Color-Color}. The distribution of all MSX sources in the region detected in the 4 MSX bands (1833 sources) is shown as small black dots whereas the 167 sources with an ISOGAL counterpart within 8$\arcsec$ are shown as black boxes. The 656 MSX sources with $F_{\rm E}/F_{\rm D}$~$\geq$~2 without an ISOGAL counterpart within 8$\arcsec$ are marked with grey circles. As seen in Fig. \ref{MSX-Color-Color}, almost half of the MSX sources in the CMZ appear in a `clump', around [8]--[15]~$\sim$~2 and [15]--[21]~$\sim$~0.5. Only $\sim$~20 of our ISO--MSX sources are located in this clump, and could actually be `normal' field objects. On the opposite, ~80\% of the ISO-MSX objects have a F$_{\rm E}$/F$_{\rm D}$ ratio above 2. Therefore, we conclude that our candidate young massive objects, selected on the basis of their ISO magnitudes, show peculiar MSX colors, and indeed trace a specific population.

\subsubsection{Interstellar extinction}

Interstellar extinction in the direction of the CMZ can exceed 30~mag in the $V$ band. Since we want to estimate the bolometric luminosities of our sources from their flux densities at 15~$\mu$m, we have to correct them for interstellar reddening for all sources in the CMZ sample.

We used an extinction map, computed by \citet{Schultheis2009}, which is based on the combination of photometric data from the 2MASS and IRAC/GLIMPSE surveys. We derived a value for $A_{\rm V}$ toward each object in our CMZ source list as an average of all data points within a radius of 0.5$\arcmin$ of the object position. For CMZ sources that are not located within the area of this extinction map, we used an older extinction map \citep{Schultheis1999} which is based on photometric data from the DENIS survey.

Furthermore, we extracted $\frac{A_{\rm 15}}{A_{\rm K_{\rm S}}}$ conversion factors from the results of \citet{Jiang2005} who give a value for each ISOGAL field. For those MSX sources that are not located in an ISOGAL field, an average value of 0.4 was set as the$\frac{A_{\rm 15}}{A_{\rm K_{\rm S}}}$ conversion factor.
The $\frac{A_{\rm K_{\rm S}}}{A_{\rm V}}$ conversion factor was obtained from the results of \citet{Nishiyama2008} ($\frac{A_{\rm K_{\rm S}}}{A_{\rm V}}$~=~0.062). This value was derived specifically for the Galactic Center environment.

The corrected magnitudes at 15~$\mu$m of the ISOGAL sources were then converted into flux densities.
The MSX flux density values were corrected by translating the flux density into magnitude values, subtracting $A_{\rm 15}$ and then retranslating the corrected magnitude values into flux densities.

\subsubsection{Bolometric luminosities}

In order to derive a conversion factor between the flux density F$_{\rm 15}$ and the luminosity $L_{\rm bol}$, we fitted the Spitzer/IRS spectra of the 49 young objects in our test sample with the SED fitting tool of \citet{Robitaille2007}. This tool contains 20000 YSO models, covering a wavelength range of 0.01 to 5000~$\mu$m, calculated at 10 different viewing angles and 50 circular apertures, which results in a total number of 10 million SEDs \citep{Robitaille2006}. The models cover a large stellar mass range of 0.1 to 50 M$_{\sun}$ and all evolutionary stages from pre-stellar to pre-main sequence. During the fitting, the tool applies all the models to the entered data, leaving the distance and the external foreground extinction $A_{\rm V}$ as free parameters within a user specified range. The result is a list of 10000 fits that are characterized by a $\chi^2$ value. Since the parameter space is sampled too sparsely to allow an accurate determination of the $\chi^2$ surface minima and, therefore, the appropriate confidence intervals, a specified $\chi^2$ criterion for "good" fits has to be selected. We defined the following criterion: $\chi^2$~--~$\chi_{\rm best}^2$~$\leq$~3~$\cdot$~$n_{\rm points}$ \citep{Robitaille2007} with $\chi_{\rm best}^2$ being the $\chi^2$ value of the best fitting model and $n_{\rm points}$ the number of data points used in the fitting process. Although this selection criterion is kind of arbitrary, a stricter criterion would imply an over-interpretation of the fitting results.

Since the distance and the external foreground extinction serve as free parameters, a distance range and a range for the visual extinction $A_{\rm V}$ have to be specified. We assumed that all CMZ sources are located at the Galactic Center distance \citep[8.0 kpc,][]{Reid2009}. Accordingly, we chose the fitted distance range to be 7--9~kpc. We set $A_{\rm V}$ to be 15 to 50~mag, corresponding to typical maximal and minimal values of $A_{\rm V}$ in the area of the 49 sources, extracted from the extinction map of \citet{Schultheis2009}.

For the fitting, we selected between six and eleven data points per spectrum, spread between 5 and 35~$\mu$m, to well reflect the underlying continuum. Since the SED tool does not model emission lines, PAH or ice features, we avoided the selection of data points in thus affected regions of the spectrum. Since the fitting tool permits to enter data taken with different apertures we can select the data points from the spectrum without correcting for the flux jumps that are visible in some of the spectra. For the size of the apertures, we chose the widths of the SL and LL modules.

We selected the models following the above defined criterion and calculated the bolometric flux densities using the luminosity values and the distance values of the different models. Then, we reconverted them into luminosity values, using a single distance of 8.0~kpc for all models. We calculated the weighted average and standard deviation of the bolometric luminosities for each source with the inverse of the $\chi^2$ values as the weights.

\begin{figure*}
	\centering
		\subfloat{\includegraphics[angle=270,width=9cm]{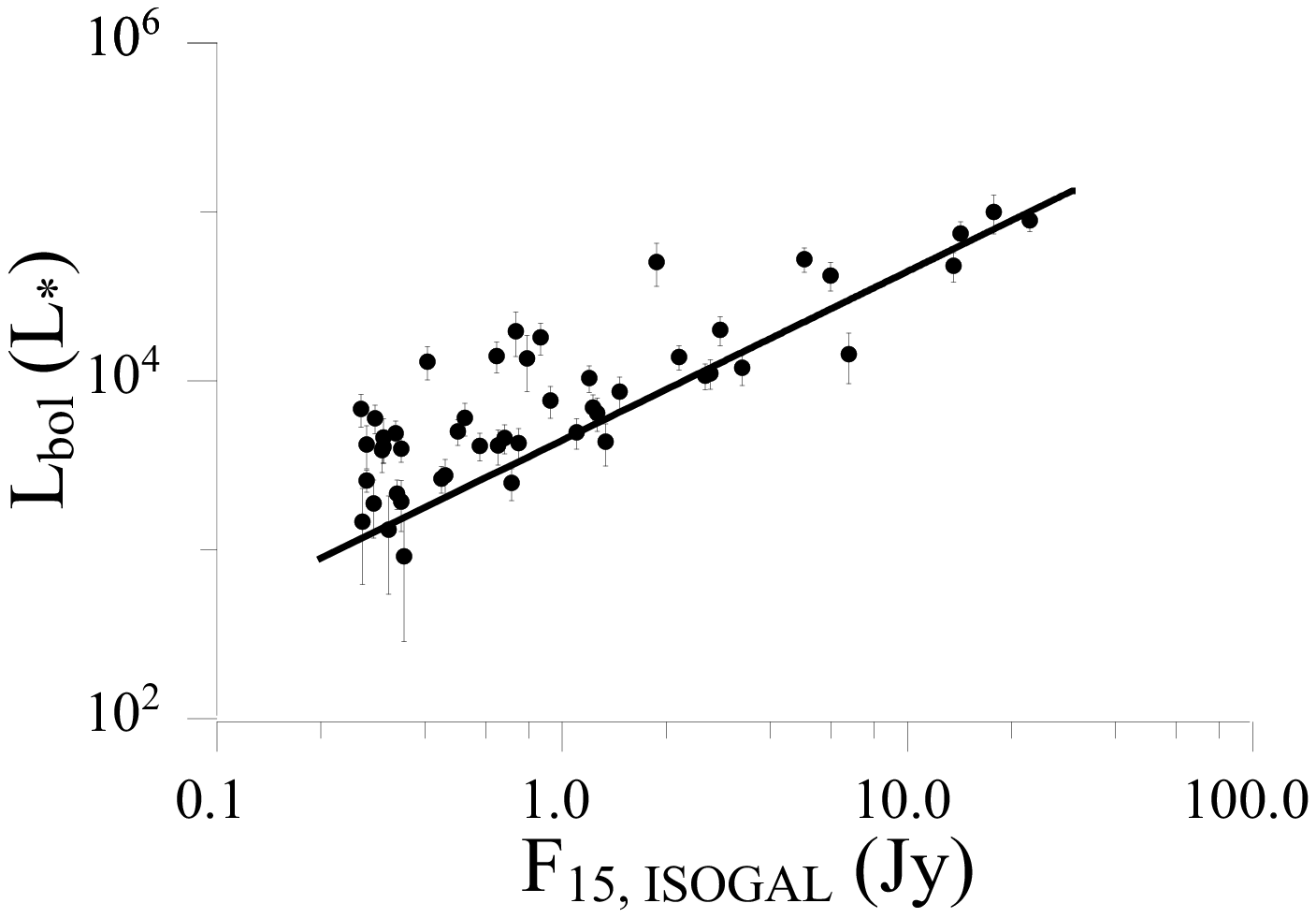}}
	\subfloat{\includegraphics[angle=270,width=9cm]{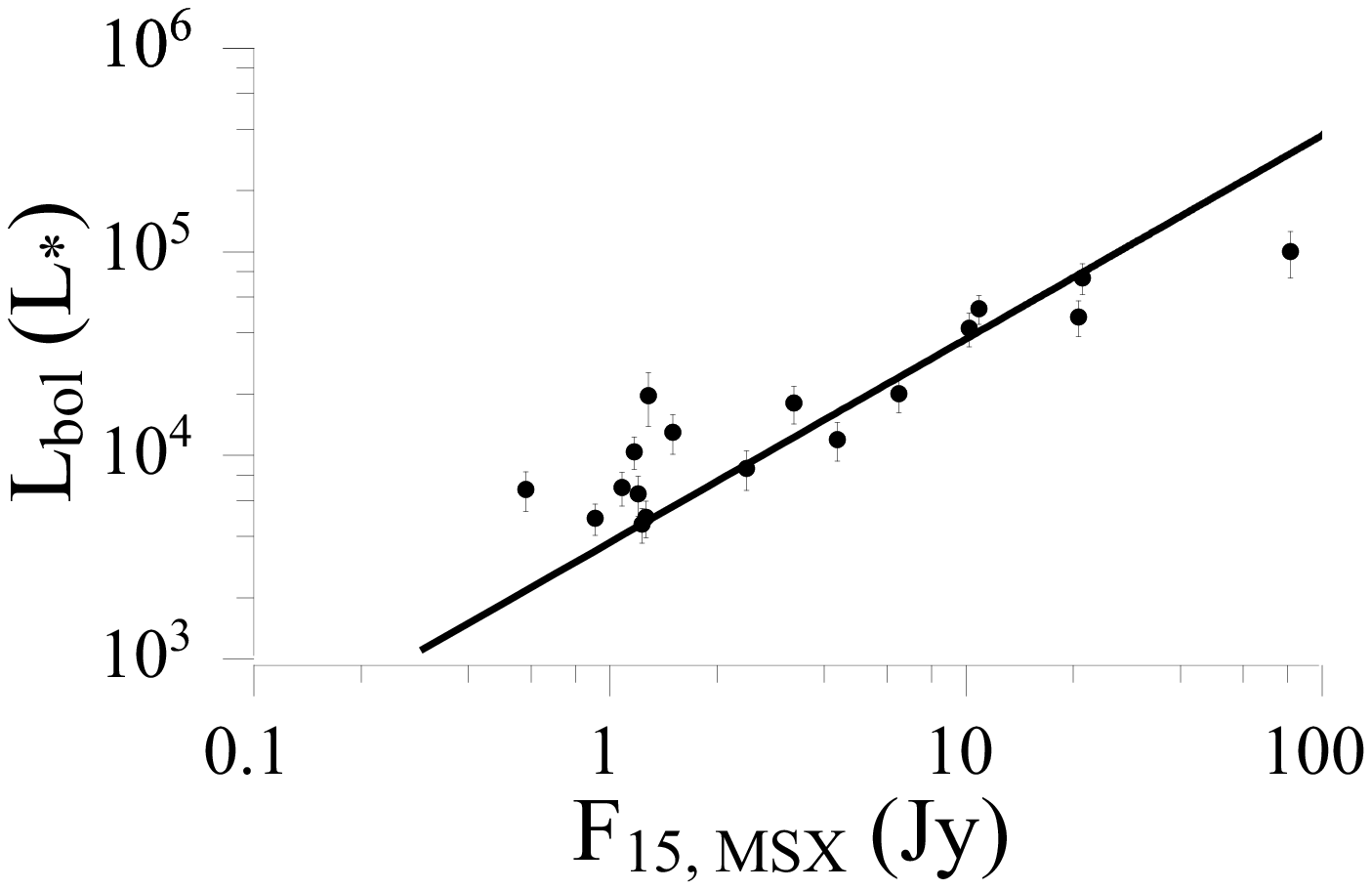}}\\	
	\caption{The images show the relation between the bolometric luminosity and the extinction corrected ISOGAL (left) and MSX (right) flux densities at 15~$\mu$m for the young objects in our test sample. The error bars correspond to the standard deviations of the bolometric luminosity values. The slopes of the fitted lines are (4447~$\pm$~106) L$_{\sun}$/Jy (left) and (3736~$\pm$~209) L$_{\sun}$/Jy (right), respectively. }
	\label{L_F_ISO_MSX}
\end{figure*}

In the next step, we plotted the obtained bolometric luminosities of the 49 ISOGAL sources against their corrected ISOGAL flux densities at 15~$\mu$m (left image in Fig. \ref{L_F_ISO_MSX}). The standard deviations served as error bar values. A linear fit to the data provided the conversion factor $(\frac{L_{\rm bol}}{F_{\rm 15}})_{\rm ISOGAL}$~=~(4447~$\pm$~106)~L$_{\sun}$/Jy. For the 18 ISOGAL sources with an MSX counterpart within 8$\arcsec$ we plotted their bolometric luminosities against the corrected MSX flux densities at 15~$\mu$m (right image in Fig. \ref{L_F_ISO_MSX}). The slope of the fitted line is  $(\frac{L_{\rm bol}}{F_{\rm 15}})_{\rm MSX}$~=~(3736~$\pm$~209)~L$_{\sun}$/Jy.

Afterwards, we calculated the bolometric luminosities for all sources in our CMZ sample from their $F_{\rm 15}$ flux densities. The luminosity values range from 0.5~$\cdot$~10$^3$ to $\sim$~10$^6$~L$_{\sun}$, with a total of $\sim$~2.5~$\cdot$~10$^7$~L$_{\sun}$.

\subsubsection{Masses}

\begin{figure}
	\centering
		\includegraphics[angle=270, width=9cm]{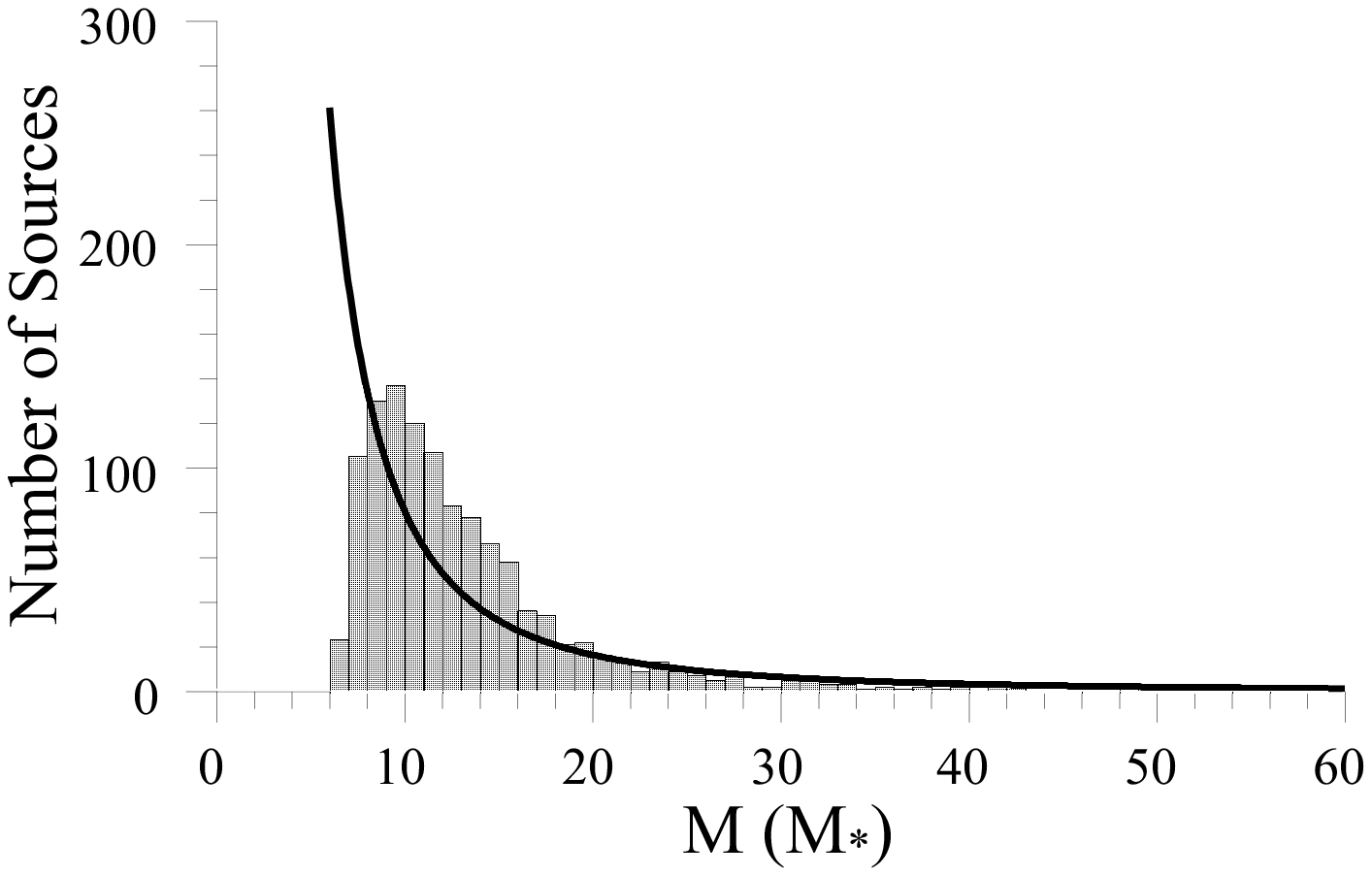}
		\caption{The image shows the distribution of the source masses in our CMZ sample. The line represents the Kroupa IMF $\xi$(M)~=~$\xi_{0,1}(M)$~$\cdot$~M$^{-2.3}$. The histogram clearly shows that low-mass and very massive objects are missing in our CMZ sample due to our selection criterion [15] < 5.25 mag and saturation effects, respectively.}
	\label{Histo_M}
\end{figure}

It is commonly accepted that hydrogen burning in high-mass stars sets in while the star is still accreting material from the circumstellar birth cloud. At this phase, the surrounding material is heated and eventually ionized by the UV radiation of the star. Based on our ISOGAL selection criteria, the sources in our CMZ sample are selected to have bright mid-infrared emission from warm dust and hence are still highly embedded in their birth clouds but already contain a central heating source. We can therefore assume that most of the sources in our CMZ sample are likely to be intermediate- to high-mass stars which have already started hydrogen burning and thus are already on the zero age main sequence (ZAMS). Assuming that the bolometric luminosity of each source is that of a single ZAMS star, mass estimates for all CMZ sample sources have been derived, using the relation \citep[e.g.][]{Eddington1926,Griffiths1988,Bhattacharya2008}
	\[\frac{L}{L_{\sun}} \approx (\frac{M}{M_{\sun}})^{3.5}.
\]

The distribution of the masses, which ranges from 6 to 58~M$_{\sun}$, is roughly consistent with a Salpeter or Kroupa initial mass function \citep[IMF,][]{Salpeter1955, Kroupa2001}. However, the histogram (see Fig. \ref{Histo_M}) clearly shows that low mass objects are missing from our CMZ sample which is due to the selection criterion [15] < 5.25 mag, corresponding to a lower mass limit of ~ 6.3 M$_{\sun}$. While most of the total luminosity in a young population comes from high mass stars, most of the total mass is contained in the high number of low mass stars. Since we are only selecting the most massive objects (the most luminous ones), we miss the stars with lower masses. In order to obtain a complete picture of the mass distribution in the CMZ, we will extrapolate the IMF to lower masses, following \citep{Kroupa2001}

\[\xi(M) = \xi_{0,1}(M) M^{-2.3} \ \textnormal{for} \ 0.5 M_{\sun} \leq M \leq 120 M_{\sun}\] 
and 
\[\xi(M) = \xi_{0,2}(M) M^{-1.3} \ \textnormal{for} \ 0.08 M_{\sun} \leq M \leq 0.5 M_{\sun}.\]
and 
\[\xi(M) = \xi_{0,3}(M) M^{-0.3} \ \textnormal{for} \ 0.01 M_{\sun} \leq M \leq 0.08 M_{\sun}.\]

We fitted the histogram with a curve of the form y~=~$\xi_{0,1}$(M)~$\cdot$~x$^{-2.3}$ and obtained $\xi_{0,1}(M)$~=~16019~M$_{\sun}^{1.3}$ (black line in Fig. \ref{Histo_M}). In order to obtain a continous IMF, we derived $\xi_{0,2}(M)$ from $\xi_{0,1}(M)$ at M~=~0.5~M$_{\sun}$ and obtained $\xi_{0,2}(M)$~=~32038~M$_{\sun}^{0.3}$ and $\xi_{0,3}(M)$ from $\xi_{0,2}(M)$ at M~=~0.08~M$_{\sun}$ and obtained $\xi_{0,3}(M)$~=~400475~M$_{\sun}^{-0.7}$ The total mass of young objects in the CMZ can then be derived with M$_{\rm tot}~=~\int~M~\xi(M)~dM$. In the mass range 0.01--120~M$_{\sun}$, the total mass is $\sim$~77000~M$_{\sun}$.

The histogram can roughly be divided into three mass ranges: 0 -- 8 M$_{\sun}$, 8--20 M$_{\sun}$ and 20 -- 60 M$_{\sun}$. As explained above, low-mass objects are missing in the lower mass range due to our selection criterion [15] < 5.25 mag. In the second mass range, the number of sources is underestimated by the fitted IMF by roughly 35\% compared to the number of sources in our CMZ sample. Confusion due to the high population in the CMZ, saturation effects of the detectors as well as rareness limit the detection of very luminous, and therefore massive, sources, which would explain the lack of objects in the upper mass range. As explained in Part I of this publication, we expect a contamination of our ISOGAL-CMZ sample with late-type evolved objects with a rate of 25\% - 35\%. If we assume that the MSX-CMZ sample is also contaminated with more evolved objects with a similar rate, we estimate the total contamination rate of our CMZ sample to be roughly 30\%. Thus, we expect to have $\sim$ 30\% more sources in our CMZ sample over the whole mass range than predicted by the fitted IMF. This effect is not noticeable in the first and third mass range due to the number of missing objects. However, in the second mass range the larger number of sources in our sample compared to the number of sources predicted by the fitted IMF can be explained with the contamination of our sample with late-type evolved objects if we assume that our sample is complete in this mass range.

\subsubsection{Stellar clusters and the Sagittarius star forming regions in the CMZ}
Several regions in the CMZ are known as active star forming sites such as the Sagittarius star forming regions (Sgr B2+B1, Sgr C, Sgr D, Sgr E) and the three massive stellar clusters that are close to the Galactic Center (Arches, Quintuplet, Central cluster). To obtain an estimate of the number of sources in our CMZ sample that are located in these active star forming regions and therefore the influence of these sites on the star formation history in the CMZ, we counted the number of ISOGAL sources and MSX sources without an ISOGAL counterpart within the area of each of these star forming regions. In Sgr B2 and B1 we found 35 ISOGAL and 32 MSX sources with a total mass of 930~M$_{\sun}$. We counted 6 ISOGAL and 6 MSX sources in Sgr C with a total mass of 201~M$_{\sun}$. In Sgr D, we detected only one MSX source with a mass of 42~M$_{\sun}$. Neither ISOGAL nor MSX sources were detected in Sgr E as well as in the three stellar clusters. These regions are probably more evolved than our CMZ sample sources since they are several million years old and their young stars are not anymore deeply embedded in their natal cloud which makes them appear less red. Thus, our sample traces more recent star formation activity.

\subsubsection{Average star formation rate}

In order to appear bright at mid-infrared wavelengths, the sources in our CMZ sample still have to be deeply embedded in the dust of their birth clouds. Following \citet{Wood1989}, this phase lasts about 10\% of the lifetime of an O-type or early B-type star, which corresponds roughly to $\sim$~1~Myr for a typical B0 star. Assuming that all objects in our CMZ sample have formed over an average time span of 1 Myr, we can convert the total mass of young objects ($\sim$~77000~M${\sun}$) to an average star formation rate of 0.08~M$_{\sun}$~yr$^{-1}$.

Previous studies \citep{Crocker2010, Yusef-Zadeh2009} published a range of values for the star formation rate in the CMZ from 0.08--0.15~M$_{\sun}$~yr$^{-1}$. \citet{An2011} make a rough estimate of the star formation rate based on their sources which are in common with the Yusef-Zadeh et al. source sample, deriving a value of 0.07~M$_{\sun}$~yr$^{-1}$.
Although our calculations are based on a number of simplifying assumptions (e.g. linear relation between F$_{\rm 15}$ and $L_{\rm bol}$, transformation of the bolometric luminosities into masses of single ZAMS stars, ...), we can state that our result is consistent with other independent studies of the star formation rate.

\section{Summary}

The goal of this study was to test and revise ISOGAL selection criteria on a test sample of 68 unknown ISOGAL sources and to apply these criteria to all ISOGAL point sources in the CMZ in order to find young objects and derive the average star formation rate over the last 1 Myr. We obtained Spitzer/IRS observations for a test sample of 68 ISOGAL sources and assembled spectra in the wavelength range 5--38~$\mu$m for 57 of these sources. In dependence of the detection of PAH or forbidden fine structure line emission and the slope of the spectrum, we classified the sources as young or late-type evolved objects. 

After revising the ISOGAL selection criteria, which are based on the ISOGAL [7]--[15] color and the spatial extent parameter $\sigma_{15}$, we applied them to all ISOGAL sources in the CMZ ($\pm$~1.5$\degr~\times~\pm$~0.5$\degr$ around the Galactic Center), selecting 485 sources. Furthermore, we added 656 MSX sources in the CMZ which fulfilled the criterion $F_{\rm E}/F_{\rm D}$~$\geq$~2 ($F_{\rm D}$, $F_{\rm E}$ flux densities in band D (15~$\mu$m) and E (21~$\mu$m)). 

After fitting the SEDs of the 47 young objects in our test sample, we obtained bolometric luminosity values and plotted them separately against the extinction-corrected ISOGAL and MSX flux densities at 15~$\mu$m of the sources. We obtained $\frac{L_{\rm bol}}{F_{15}}$ conversion factors and applied them to our CMZ sample. Assuming that the bolometric luminosity of each source corresponds to the luminosity of a ZAMS star, we calculated the mass for each source. The distribution of the masses roughly follows a Salpeter or Kroupa IMF but low-mass and very massive objects are missing from the CMZ sample. The total mass of young objects in the CMZ ($\sim$~77000~M$_{\sun}$) was calculated from a Kroupa IMF between 0.01 and 120~M$_{\sun}$. Since the detection as bright mid-infrared sources requires that the sources are still deeply embedded in their dust cocoon, the sources still have to be very young. Assuming that this total mass in young objects has formed over the last 1~Myr, we obtained a star formation rate of 0.08~M$_{\sun}$~yr$^{-1}$ for the CMZ which is consistent with previous studies.

\bibliographystyle{aa}
\bibliography{ReferencesPaper}

\end{document}